\documentstyle[12pt]{article}
\def\ftoday{{\sl  \number\day \space\ifcase\month 
\or Janvier\or F\'evrier\or Mars\or avril\or Mai
\or Juin\or Juillet\or Ao\^ut\or Septembre\or Octobre
\or Novembre \or D\'ecembre\fi
\space  \number\year}}    
\newcommand{\journal}[4]{{\em #1~}#2\,(19#3)\,#4;}

\newcommand{\ijmp}{\journal {Int. J. Mod. Phys.}}

\newcommand{\pr}{\journal {Phys. Rev.}}

\newcommand{\cmp}{\journal {Commun. Math. Phys.}}

\newcommand{\cqg}{\journal {Class. Quantum Grav.}}

\newcommand{\np}{\journal {Nucl. Phys.}}
\newcommand{\pl}{\journal {Phys. Lett.}}

\newcommand{\prep}{\journal {Phys. Rep.}}

\newcommand{\annp}{\journal {Ann. Phys. (N.Y.)}}

\makeatletter
\@addtoreset{equation}{section}
\makeatother

\newcommand{\es}{\\[3mm]}

\renewcommand{\a}{\alpha}
\renewcommand{\b}{\beta}
\newcommand{\g}{\gamma}           \newcommand{\G}{\Gamma}
\renewcommand{\d}{\delta}         \newcommand{\D}{\Delta}
\newcommand{\e}{\varepsilon}
\newcommand{\k}{\kappa}
\newcommand{\la}{\lambda}        
\newcommand{\m}{\mu}
\newcommand{\n}{\nu}
\newcommand{\om}{\omega}         \newcommand{\OM}{\Omega}
\newcommand{\p}{\psi}              
\newcommand{\s}{\sigma}           \renewcommand{\S}{\Sigma}
\newcommand{\th}{\theta}         
\newcommand{\f}{{\phi}}           \newcommand{\F}{{\Phi}}
\newcommand{\vf}{{\varphi}}
\newcommand{\GG}{{\cal G}}

\newcommand{\xint}{\dint d^4x\;}
\newcommand{\sla}{\raise.15ex\hbox{$/$}\kern -.57em} 
\newcommand{\Sla}{\raise.15ex\hbox{$/$}\kern -.70em}
\def\h{\hbar}

\def\LP{\displaystyle{\Biggl(}}
\def\RP{\displaystyle{\Biggr)}}
\newcommand{\lp}{\left(}\newcommand{\rp}{\right)}
\newcommand{\lc}{\left[}\newcommand{\rc}{\right]}
\newcommand{\lac}{\left\{}\newcommand{\rac}{\right\}}

\newcommand{\complex}{{\kern .1em {\raise .47ex
\hbox {$\scriptscriptstyle |$}}
    \kern -.4em {\rm C}}}
\newcommand{\real}{{{\rm I} \kern -.19em {\rm R}}}
\newcommand{\rational}{{\kern .1em {\raise .47ex
\hbox{$\scripscriptstyle |$}}
    \kern -.35em {\rm Q}}}
\renewcommand{\natural}{{\vrule height 1.6ex width
.05em depth 0ex \kern -.35em {\rm N}}}

\newcommand{\tr}{{\rm {Tr} \,}}
\newcommand{\half}{\dfrac{1}{2}}
\newcommand{\pa}{\partial}

\newcommand{\dpad}[2]{{\displaystyle{\frac{\partial #1}{\partial #2}}}}
\newcommand{\dfud}[2]{{\displaystyle{\frac{\delta #1}{\delta #2}}}}
\newcommand{\dfrac}[2]{{\displaystyle{\frac{#1}{#2}}}}
\newcommand{\dsum}[2]{\displaystyle{\sum_{#1}^{#2}}}   
\newcommand{\dint}{\displaystyle{\int}}

\newcommand{\twiddle}{\lower.9ex\rlap{$\kern -.1em\scriptstyle\sim$}}

\newcommand{\bra}[1]{\left\langle {#1}\right|}
\newcommand{\ket}[1]{\left| {#1}\right\rangle}

\newcommand{\equ}[1]{(\ref{#1})}
\newcommand{\eq}{\begin{equation}}
\newcommand{\eqn}[1]{\label{#1}\end{equation}}
\newcommand{\eea}{\end{eqnarray}}
\newcommand{\eqa}{\begin{eqnarray}}
\newcommand{\eqan}[1]{\label{#1}\end{eqnarray}}
\newcommand{\ba}{\begin{array}}
\newcommand{\ea}{\end{array}}
\newcommand{\eqac}{\begin{equation}\begin{array}{rcl}}
\newcommand{\eqacn}[1]{\end{array}\label{#1}\end{equation}}

\def\smab{\s^{\hspace{1mm}\m}_{\a\bd}}
\def\esmab{\e^\a\smab\eb^{\bd}}
\newcommand{\lb}{\bar\lambda}

\newcommand{\biblio}[1]{{\it Bibliography:} \cite{#1}\hfill\break}
\newcommand{\oilbib}{\vspace{2mm}
\noindent}
\newcommand{\point}[1]{\vspace{3mm}
\noindent{\bf #1}}
\newcommand{\proposition}{\vspace{1mm}
\noindent{%\large
{\bf Proposition. }}}
\newcommand{\proof}{\vspace{3mm}
\noindent{%\large
{\bf Proof: }}}
\newcommand{\foorp}{ {\hfill$\Box$}  \vspace{2mm}
}
\newcommand{\remark}{\vspace{3mm}
\noindent{%\large
{\bf Remark. }}}
\newcommand{\kramer}{\vspace{2mm}
\noindent}
\newcommand{\remarks}{\vspace{3mm}
\noindent{%\large
{\bf Remarks.}}\begin{enumerate} }
\newcommand{\skramer}{\end{enumerate}%\vspace{2mm}
\noindent}
\newcommand{\bi}{\begin{itemize}} \newcommand{\ei}{\end{itemize}}

\newcommand{\bint}[2]{{\dint_{\kern -.4em #1}^{#2}}}

\newcommand{\Bsurl}[1]{ 
{\raise 5ex \hbox{$\overline{{\raise -5ex \hbox{$#1$} }}$} }
}
\newcommand{\bsurl}[1]{ 
{\raise 3ex \hbox{$\overline{{\raise -3ex \hbox{$#1$} }}$} }
}
\newcommand{\surl}[1]{ 
{\raise 2ex \hbox{$\overline{{\raise -2ex \hbox{$#1$} }}$} }
}
\newcommand{\ad}{{\dot\a}}  \newcommand{\bd}{{\dot\b}}

\newcommand{\QB}{{\bar{Q}}}

\newcommand{\AB}{{\bar{A}}}
\newcommand{\FB}{{\bar{F}}}

\newcommand{\cb}{{\bar{c}}}
\newcommand{\db}{{\bar\d}}
\newcommand{\eb}{{\bar\e}}
\newcommand{\sib}{{\bar\s}}

\newcommand{\psb}{{\bar\p}} \newcommand{\chib}{{\bar\chi}}
\newcommand{\smuaad}{\s^\m_{\a\ad}}
\newcommand{\sbmuaad}{{\bar\s}_\m^{\ad\a}}

\begin{document}
%***********************************************************************
%\thispagestyle{empty}
\begin{titlepage}

{\hfill \parbox{45mm}{{\large UFES--DF--OP97/1\\[2mm]
                             hep-th/yymmxxx}}}
\vspace{15mm}

\begin{center}{\huge{\bf Introduction to \\[2mm] 
                        Supersymmetric Gauge Theories}}
\\[12mm]
{\Large  Olivier Piguet}\footnote{On leave of
absence from:
{\it D\'epartement de Physique Th\'eorique --
     Universit\'e de Gen\`eve, 24 quai E. Ansermet -- CH-1211 Gen\`eve
     4, Switzerland.}}$^,$\footnote{Supported in part by the 
     Swiss National Science Foundation and the Brazilian National Research 
     Council (CNPq)}\\[1cm]
{\small Departamento de F\'\i sica, 
   Universidade Federal do  Esp\'\i rito Santo\\
   BR-29060-900 Vit\'oria, E.S.,  Brasil}

%%%%%%%%%%%%%%%%%%%%%%%%%%%%%%%%%%%%%%%%%%%%%%%%%%%%%%%%%%%%%%%%%%%%%
\vspace{1cm}

{Lectures given at the ``First School on Field Theory and
Gravitation'', Universidade Federal do Esp\'\i rito Santo, 
Vit\'oria, Brazil, 
April 1997}\\[2cm]
\end{center}
\begin{abstract}
In these lectures I present a basic introduction 
to supersymmetry, especially
to $N=1$ supersymmetric gauge theories and their renormalization,
in the Wess-Zumino gauge. I also discuss the
various ways supersymmetry may be broken in order to account for the
lack of exact supersymmetry in the actual world of elementary particles.
\end{abstract}

\end{titlepage}
%------------------------------------------------------------------

%******************************************************
%\tableofcontents
%******************************************************

%***************************************************************************
\section*{Introduction}\label{introduction}
This course intends to be an introduction to supersymmetry, and in
particular to $N=1$ supersymmetric gauge theories. 
I shall present here the formalism in components,  in the so-called 
Wess-Zumino gauge (WZ-gauge). Whereas many 
books~\cite{gates-book,bagger-book,srivatsana,west-book,buchbinder,ps-book}
and reviews~\cite{sohnius,lykken,reviews,susy96} 
on the superspace formalism are available, few are especially devoted to 
the formalism in the WZ-gauge. There are however two good reasons for studying the 
latter, despite of the
simplicity and of the beauty of the superspace formalism. The first one
is that in the physically interesting case where supersymmetry is 
broken, e.g. by the presence of nonsupersymmetric masses, a good part of
the advantages of superspace is lost. The second motivation is that
there is no simple superspace formalism for extended ($N>1$) 
supersymmetry\footnote{The 
superspace of extended supersymmetry is  the ``harmonic 
superspace''~\cite{harm-s-space}. It 
is not as simple to handle with as the $N=1$ superspace.}: 
these lectures may thus provide also a basis for a further
study including these extended theories.

The plan of the lectures is as follows. 
I shall begin, in Section~\ref{susy-non-rel},
with the description of a simple 
nonrelativistic supersymmetric
quantum model, in order to familiarize ourselves with some of the
features pecular to supersymmetry.
Then, after some
generalities about $N$-extended supersymmetry and its representations 
in the spaces of one-particle states, in Section~\ref{rel-susy-gen},
I shall introduce the $N=1$ chiral supermultiplet, 
in the component formalism, in the fist part of 
Section~\ref{superm.-de-champs}, 
and apply it to a simple model. 
In the second part of Section~\ref{superm.-de-champs}, 
I shall introduce the gauge supermultiplet in the
Wess-Zumino gauge and describe the general supersymmetric gauge invariant
action. Gauge fixing, BRS invariance and renormalization 
of supersymmetric gauge theories in the Wess-Zumino gauge will be
discusssed in Section~\ref{renormalisation}. 
In Section~\ref{susy-brisee}, 
I shall finally give some comments on the different ways
one may break supersymmetry (explicitly or spontaneously)
in order to account for the absence of exact
superymmetry in nature, and present a case of spontaneous breakdown of
supersymmetry accompagnied by the phenomenon of radiative mass generation.

%************************************************************
\section{Nonrelativistic Supersymmetry}\label{susy-non-rel}
In order to get familiarized with the concept of supersymmetry,
let us introduce it  first in the framework of nonrelativistic quantum
mechanics. Although various applications may be found in the 
literature~\cite{susy-harm-osc}, I shall restrict myself
to the simple  model
of the {\it supersymmetric harmonic 
oscillator}. It consists of a bosonic and of a
fermionic part.
%**************************************************************
\subsection{The Bosonic Oscillator}
The bosonic part is the usual one-dimensional harmonic oscillator, 
defined by the Hamiltonian 
\eq
H_B = \half(p^2+\om_B^2)\ .
\eqn{bos-hamilt}
In terms of the annihilation and creation operators
\[
 a = \dfrac{1}{\sqrt{2\om_B}}(p-i\om_B x)\ ,\quad
  a^+ = \dfrac{1}{\sqrt{2\om_B}}(p+i\om_B x)
\]
the Heisenberg commutation relation $[p,x]=-i$ writes\footnote{We take 
$\h=1$. The superscript $^+$ means Hermitean conjugation.}
\eq
[a,a^+] = 1\ ,
\eqn{bos-com-rel}
and the Hamiltonian takes the form
\eq
H_B = \dfrac{\om_B}{2}(a^+a + aa^+) =  \om_B\lp N_B+\half \rp \ ,
\eqn{bos-hamilt'}
where $N_B=a^+a$ is a ``counting operator'', with eigenvalues 
$0,1,2,\cdots$.
The energy spectrum is thus given by
\eq
E_B= \om_B\lp n_B + \half\rp\ ,\quad n_B = 0,1,2,\cdots\ .
\eqn{bos-spectrum}
%****************************************************************
\subsection{The Fermionic Oscillator}
One can define in an analogous way a fermionic harmonic oscillator, by
introducing fermionic annihilation and creation operators $b$, $b^+$ 
obeying the anticommutation rules
\eq
\{b,b^+\} = 1\ ,\quad \{b,b\} = \{b^+,b^+\} = 0\ ,
\eqn{ferm-anticom}
and a Hamiltonian
\eq
H_F = \dfrac{\om_F}{2}(b^+b - bb^+) =  \om_F\lp N_F-\half \rp \ ,
\eqn{ferm-hamilt'}
where $N_F=b^+b$ is the fermionic counting operator, 
with eigenvalues $0,1$.
The fermionic energy spectrum thus is
\eq
E_F= \om_F\lp n_F - \half\rp\ ,\quad n_F = 0,1,2,\cdots\ .
\eqn{ferm-spectrum}
An explicit representation of the fermionic operators, preserving the
anticommutation rules \equ{ferm-anticom}, is given in terms of Pauli
matrices by\footnote{$\s_\pm=\s_1\pm i\s_2$, $\ [\s_+,\s_-]= \s_3$.}
\[
b  = \s_- =  \lp\ba{cc} 0\ &0\\1 &0 \ea\rp\ ,\quad
b^+= \s_+ =  \lp\ba{cc} 0\ &1\\0 &0 \ea\rp\ .
\]
The fermionic Hamiltonian then takes the form of the Hamiltonian for a
system consisting of one spin $1/2$:
\eq
H_F = \dfrac{\om_F}{2}\s_3\ .
\eqn{ferm-hamilt}
%****************************************************************
\subsection{The Supersymmetric Oscillator}
The supersymmetric oscillator is obtained by combining one bosonic and
one fer\-mion\-ic oscillators \equ{bos-hamilt'}, \equ{ferm-hamilt'}
with the same frequency
$\om=\om_B=\om_F$. Its Hamiltonian is given by
\eq\ba{l}
H = H_B + H_F  = \dfrac{\om}{2}(a^+a + aa^+) + 
\dfrac{\om}{2}(b^+b - bb^+)\\[4mm]
\phantom{H = H_B + H } = \half p^2 + \half\om^2 x^2 + \half \om \s_3 \ .
\ea\eqn{susy-hamilt}
The spectrum is obtained as the sum of \equ{bos-spectrum} and 
\equ{ferm-spectrum}:
\eq
E_{n_B,n_F} = \om\lp n_B+n_F \rp\ ,\quad n_B=0,1,2,\cdots\ ,\ 
n_F=0,1\ .
\eqn{susy-spectrum}
It corresponds to a degenerate groundstate
\[
\p_{0,0}\ ,\quad\mbox{of energy }E_{0,0}=0\ ,
\]
and of a sequence of doubly degenerated of states
\[
\p_{n,0}\ \mbox{and}\ \p_{n-1,1}\ ,\quad\mbox{of energy }
  E_{n,0} = E_{n-1,1} = n\om\ ,\quad n=1,2,\cdots\ .
\]
As one may expect, the degeneracy of the excited states is due to a
symmetry. Let us indeed introduce two Hermitean {\it supercharges}
\eq
Q_1 = \sqrt{\om} \lp a^+b+ab^+ \rp\ ,\quad 
Q_2 = \sqrt{\om} i\lp a^+b-ab^+ \rp\ .
\eqn{s-charges}
One easily sees that, together with the Hamiltonian \equ{susy-hamilt},
they obey the {\it sup\-er\-al\-geb\-ra\footnote{With the notation
$\{A,B\}=AB+BA\ ,\quad [A,B]=AB-BA$.}}
\eq
\lac Q_i,Q_j \rac = 2\d_{ij} H\ ,\quad \lc Q_i, H \rc = 0\ .
\eqn{s-algebra}
In particular, we have
\eq
H=(Q_1)^2=(Q_2)^2\ .
\eqn{H=Qsquare}
The second of eqs. \equ{s-algebra}, namely the commutativity of the
supercharges with the Hamiltonian, follows from the special case
\equ{H=Qsquare} of the first one. It characterizes them as the
generators of a symmetry, called {\it supersymmetry}.
The infinitesimal supersymmetry transformations of the fundamental variables 
$x$, $p$, $\s_1$ and $\s_2$ are easily calculated. They read:
\eq\ba{l}
\d_k x = i[Q_k,x] = \dfrac{1}{\sqrt2} \s_k\ ,\quad 
\d_k p = i[Q_k,p] = \dfrac{i\om}{\sqrt2} \e_{kl}\s_l\ , \es
\d_k \s_l =
i\{Q_k,\s_l\} = \d_{kl}\dfrac{1}{\sqrt2} p + \e_{kl}\sqrt2 \om x\ ,
\ea\eqn{susy-transf}
with $\e_{kl}=-\e_{lk}$, $\e_{12}=1$.

It is worth emphasizing three important properties of this system: the
nonnegativity of the energy, the double degeneracy of the energy levels,
and the vanishing ground state energy. These properties are
generic properties of supersymmetry. As the following proposition 
shows, they
follow merely from the superalgebra \equ{s-algebra}:
\proposition
\begin{enumerate}
\item[(i)] The energy eigenvalues are nonnegative.
\item[(ii)] The ground state energy is vanishing if and only if the
ground state is invariant under supersymmetry transformations. It is then 
a supersymmetry singlet. 
\item[(iii)] If the ground state energy is nonzero, then the ground state
is not invariant under supersymmetry. In other words, supersymmetry is
then spontaneously broken.
\end{enumerate}
\proof (i) Energy positivity follows from the equalities, valid for 
any state vector $\ket{\p}$:
\eq
\bra{\p} H \ket{\p} = \bra{\p} Q^1Q^1 \ket{\p} = \|Q^1\ket{\p}\|^2 \ge0\ .
\eqn{energy-vev}
which are a consequence of \equ{H=Qsquare}.
\noindent (ii) Let us suppose that the ground state 
$\ket{0}$ is supersymmetric: $Q^i\ket{0}=0$ for $i=1,2$. $H\ket{0}=0$ 
then follows from \equ{H=Qsquare}. Conversely, 
$\bra{0} H \ket{0}=0$ implies $Q^i\ket{0}=0$ due to \equ{energy-vev} 
and the positive definiteness of the norm.

\noindent (iii) is a corollary of (i) and (ii).
\foorp
\remark
In our exemple of the supersymmetric oscillator, we saw explicitly  
in \equ{susy-spectrum} how the bosonic and fermionic contributions to 
the ground state energy, $\om/2$ and $-\om/2$, respectively, 
exactly cancel. As point (ii) of the proposition above shows, 
this boson-fermion cancellation in the ground state energy, 
following from the 
superalgebra alone, is quite general.

%******************************************************
\section{Relativistic Supersymmetry. Generalities}\label{rel-susy-gen}
In 1967, Coleman and Mandula~\cite{col-mand} proved that there is no
nontrivial way to  unify an internal symmetry with 
the relativistic space-time
symmetry. In more precise words, any symmetry group $\GG$ containing 
the Poincar\'e group and an arbitrary internal symmetry group has
the form of a direct product
\[
\GG = \mbox{ Poincar\'e }\times\mbox{ Internal Symmetry } \ .
\]
This theorem concerns the symmetries of the $S$-matrix. It is based on
quite general hypothese such as  locality and unitarity, and on the
assumption that the symmetries are described by Lie groups. 
Coleman-Mandula's theorem has put an end to numerous tentatives of
unification of internal and space-time symmetries.
However, in 1975, Haag, {\L}opusza\'nski and Sohnius~\cite{haag-lo-so}
found a way out of this ``no-go'' theorem, through the relaxation 
of the assumption that the group of symmetries is a Lie group, whose
generators obey a Lie algebra in the form of {\it commutation} rules. 
Allowing indeed for the presence of {\it anticommutaton} rules 
as well,
they showed that the so-called Poincar\'e supersymmetry allows
for a nontrivial unification. Their result is nevertheless very
restrictive, in the sense that beyond the usual bosonic scalar 
generators -- Lie
group generators obeying commutation rules -- the only other possible
ones are fermionic  generators of spin 1/2 -- obeying anticommutation rules.
%**************************************************************
\subsection{Poincar\'e Supersymmetry Algebras}
According to the result of Haag, {\L}opusza\'nski and 
Sohnius~\cite{haag-lo-so}, 
the basis of the super-Poincar\'e algebra with $N$ supersymmetry 
generators (``$N$-supersymmetry'') consists of:
\bi
\item
Bosonic ({\it even}) Hermitean scalar generators $T_a$, 
$a=1\cdots \mbox{dim}(G)$, of some internal symmetry Lie group $G$,  
\item The ({\it even}) generators $P_\m$ and $M_{[\m\n]}$ of the 4-dimensional
Poincar\'e group.
\item
Fermionic ({\it odd}) Weyl spinor generators $Q_\a^i$, 
$\a=1,2$; $i=1,\cdots,N$, belonging to a dimension $N$ representation 
of $G$, 
and  their conjugates $\QB^\ad_i$. 
\item 
Central charges $Z^{[ij]}$, i.e. bosonic operators commuting with all the
$T_a$'s and all the $Q_\a$'s, as well as with the
Poincar\'e generators.
\ei
The $T_a$'s and $Z^{[ij]}$'s are scalars, whereas the $Q_\a^i$'s belong to the
representation $(1/2,0)$ of the Lorentz group and the $\QB^\ad_i$'s to
the conjugate representation $(0,1/2)$. The $Q_\a$'s are  
Weyl spinors, with two complex 
components\footnote{The notations and conventions are
given in the Appendix.}.  

Moreover, the general  superalgebra of $N$-extended supersymmetry, also
called the $N$-super-Poincar\'e algebra, reads
(we write only the nonvanishing (anti)com\-mut\-at\-ors):
\eq\ba{l}
[M_{\m\n},M_{\rho\s}] = -i(g_{\m\rho}M_{\n\s} - g_{\m\s}M_{\n\rho} +
                        g_{\n\s}M_{\m\rho} - g_{\n\rho}M_{\m\s} )\ ,\es
[M_{\m\n},P_\la] = i(P_\m g_{\n\la} - P_\n g_{\m\la})\ ,
\ea\eqn{alg-poincare}
\eq\ba{l}
[T_a,T_b] = i {f_{ab}}^c T_c\ ,
\ea\eqn{alg-lie}
\eq\ba{l}
\lac Q^i_\a,Q^j_\b \rac = \e_{\a\b} Z^{[ij]}\ ,\es
\lac Q^i_\a,\QB^j_\ad \rac = 2\d^{ij} \smuaad P_\m  \ ,\
\ea\eqn{alg-susy}
\eq\ba{l}
[Q^i_\a,M_{\m\n}]= \dfrac{1}{2} {(\s_{\m\n})_\a}^\b Q^i_\b\ ,\es
[Q^i_\a,T_a] = {(R_a)^i}_j Q^j_\a\ . 
\ea\eqn{alg-lie-susy}
This result is the most general one for a massive theory. 
In a massless theory, another
set of fermionic charges, $S^i_\a$ (and their conjugates), 
may be present. 
Then, in this case, the Lie group $G$ is restricted to be U($N$), 
for $N\not=4$, and
either U(4) or SU(4), for $N$=4. The superalgebra moreover 
contains all the
generators of the conformal group -- which contains the Poincar\'e group as
a subgroup: one calls it the $N$-superconformal algebra.

In these lectures we shall restrict ourselves to the case $N$=1, 
where the part of the superalgebra 
\equ{alg-poincare}-\equ{alg-lie-susy} 
which involves the spinor charges reduces 
to the original Wess-Zumino algebra 
\eq
\lac Q_\a,\QB_\ad \rac = 2 \smuaad P_\m  \ ,\quad
\lac Q_\a,Q_\b \rac = 0\ ,\quad \lac \QB_\ad,\QB_\bd \rac = 0\ ,
\eqn{alg-wz}
and
\eq\ba{l}
[Q_\a,M_{\m\n}]= \dfrac{1}{2} {(\s_{\m\n})_\a}^\b Q_\b\ ,\quad
[\QB^\ad,M_{\m\n}]= -\dfrac{1}{2} {(\sib_{\m\n})^\ad}_\bd \QB^\bd\ ,\es
[Q_\a,P_\m]= 0\ ,\quad [\QB_\ad,P_\m]= 0\ ,
\ea\eqn{com-poinc-susy}

\eq
[ Q_\a,R] = -Q_\a\ ,\quad [\QB_\ad,R] = \QB_\ad\ .
\eqn{com-r-susy} 
Here, $R$ is the infinitesimal generator of an Abelian group into
which the internal symmetry group $G$ has shrunk.
\remark
The elements of the Pauli matrices $\s^\m$, $\s_{\m\n}$ and $\bar\s_{\m\n}$
appear as the structure constants of this superalgebra.
\kramer
%************************************************************
\subsection{Representations in the Space of Single Particle 
States}
\biblio{sohnius,lykken}
\subsubsection*{Massive particles:}
Massive particles are defined as irreducible representations (irrep.)
of the Poincar\'e 
group, labeled, in the massive case,
by their mass $m$ and their spin $j$. One-particle states can be taken
as eigenstates of the four momentum 
$p=(p_\m,\,\m=0,\cdots,3)$, with $p^2=m^2$,  and of the spin component 
$j_3$, with $j_3=-j,-j+1,\cdots,j$. It
is convenient to work in the rest frame $p=(m,\bf 0)$, where the
Wess-Zumino algebra \equ{alg-wz} takes the simple form
\[
 \{Q_\a,\QB_\ad\} = 2m\d_{\a\ad}\ ,\quad
   \{Q_\a,Q_\b\} =  \{\QB_\ad,\QB_\bd\} = 0\ .
\]
We see that the operators 
\[ 
a_\a=\dfrac{1}{\sqrt{2m}}Q_\a\ ,\quad a^+_\a=\dfrac{1}{\sqrt{2m}}\QB_\ad
\]
obey the algebra of two pairs of fermionic annihilation and creation
operators: 
\[
\{a_\a, a^+_\b\} = \d_{\a\b}\ ,\quad
 \{a_\a, a_\b\} = \{a^+_\a, a^+_\b\} = 0\ ,\quad \a,\b=1,2 \ .
\]
From the form of this algebra it follows that one can construct all
its irreducible representations by acting with the creation operators 
$a^+_\a$ on a state $\ket{\Omega}$ 
-- analogous to the vacuum state in usual Fock space -- characterized by
\[
a_\a\ket{\Omega}=0\ ,\quad \a=1,2\ .
\]
The states of the irrep. defined from $\ket{\Omega}$ are thus:
\eq
\ket{\Omega}\ ,\qquad a^+_\a\ket{\Omega}\,,\ \a=1,2\ ,\qquad 
a^+_1a^+_2\ket{\Omega}=-a^+_1a^+_2\ket{\Omega} \ .
\eqn{massive-irrep}
Let us assume that $\ket{\Omega}$ is a one-particle state of 
spin $j$, with $j_3=\m$. Then, from the commutation relation
\[
[J^i,a^+_\a]= \half\s^i_\a{}^\b a^+_\b\ ,
\]
which is a special case of \equ{com-poinc-susy} for the rotation group
generators $J^i=\dfrac{1}{2}\e^{ijk}M_{jk}$, it follows that the states 
in \equ{massive-irrep} have the spin components $\m$, $\m\pm 1/2$ and $\m$,
respectively.
If one takes the $2j+1$ possible states $\OM$ with $-j\le\m\le j$, the 
multiplicities of the three ``levels'' in \equ{massive-irrep} are
$2j+1$, $2(2j+1)$ and $2j+1$, 
respectively. One observes that the numbers of bosonic and fermionic
degrees of freedom are equal, each, to $2(2j+1)$.
In the special case $j=0$, we have two spin 0 particles and one spin 1/2
particle (with $\m=\pm1/2$).
\remark
These representations, of dimension $4(2j+1)$,
are irreducible as representations of the algebra
generated by $Q_\a$, $\QB_\ad$, $P_\m$ and $M_{\m\n}$. As representations 
of the Wess-Zumino algebra \equ{alg-wz}, 
generated by $Q_\a$, $\QB_\ad$ and $P_\m$ only, they are
irreducible only for  $j=0$. 
\kramer

%*********************************************************** 
\subsubsection*{Massless particles:}
Massless one-particle states are characterized by their 3-momentum 
$\bf p$ and their
helicity $\Lambda=\bf J\cdot\bf p/|\bf p|$.

Let us consider 
states $\ket{E,\la}$ of 4-momentum $p=(E,0,0,E)$ and of helicity
eigenvalue $\la$, with $\Lambda\ket{E,\la}=\la\ket{E,\la}$,
choosing a frame where the 3-momentum is parallel to
the $z$-axis.
On such states, the $Q$, $\QB$ anticommutation rule 
\equ{alg-wz} reduces to
\[
\{ Q_\a,\QB_\ad \} = 2\lp \s^0+\s^3\rp_{\a\ad} E\ ,
\]
i.e.:
\[
\{Q_1,\QB_1\} = 4E\ ,\quad 
\{Q_1,\QB_2\} = \{Q_2,\QB_1\} = \{Q_2,\QB_2\} = 0 \ .
\]
It follows in particular, from the last anticommutation relation and
from the positivity of the operator $Q_2\QB_2$, that $Q_2=0$ on these
states. The nontrivial part of the algebra has thus the form
\[
\{a,a^+\}=1\ , \quad \{a,a\}=\{a^+,a^+\}=0\ ,
\quad\mbox{with}\quad a=\dfrac{1}{2\sqrt{E}}Q_1\ ,\quad
  a^+=\dfrac{1}{2\sqrt{E}}\QB_1\ .
\]
We have now only one pair of 
fermionic creation and annihilation operators: the
irrep. are thus two-dimensional. Starting with a state
$\ket{\OM}$ annihilated by $a$: $a\ket{\OM}=0$, we get the doublet 
\eq
\ket{\OM}\ ,\quad a^+\ket{\OM}\ .
\eqn{massless-irrep}
If $\ket{\OM}$ is a state of helicity $\la$, then, 
from the commutation relation 
\[
[J^3,a^+]= \half a^+\ ,
\]
follows that the helicity of its partner $a^+\ket{\OM}$ is $\la+1/2$.
\remark
Considering only the proper and orthochronous component of the 
Lorentz group, which leaves helicity invariant, we see that the
doublets are irreducible representations of the supergroup generated by
the whole algebra including the Lorentz generators. Of course, if parity
is included, then an irrep. is 4-dimensional, consisting of the states of
helicity $\la$, $\la+1/2$, $-\la$ and $-\la-1/2$.
\kramer

%*********************************************************** 
\subsubsection*{Generalization for $N>1$ (Massless particles):}

The generalization to extended supersymmetry,
$N>1$, is straightforward if we 
assume the absence of central charges. We shall moreover
restrict ourselves 
to the case of massless particles. With the same kinematics assumptions
as in the case $N=1$ above, the extended superalgebra reads
\[
\{Q^i_1,\QB^j_1\} = 4\d_{ij}E\ ,\quad 
\{Q^i_1,\QB^j_2\} = \{Q^i_2,\QB^j_1\} = \{Q^i_2,\QB^j_2\} = 0 \ ,
\ i,j=1,\cdots,N\ .
\]
Again, $Q^i_2$ vanishes on the helicity states $\ket{E,\la}$, 
and the nontrivial part of the algebra takes the form
\[
\{a_i,a_j^+\}=\d_{ij} , \quad \{a_i,a_j\} = \{a_i^+,a_j^+\}=0\ ,\es
\mbox{with}\quad a_i=\dfrac{1}{2\sqrt{E}}Q^i_1\ ,\quad
  a_i^+=\dfrac{1}{2\sqrt{E}}\QB^i_1\ .
\]
Starting from the state $\ket{\OM}=\ket{E,\la}$ 
obeying the conditions 
$a_i\ket{\OM}=0$, we construct the supermultiplet
\eq\ba{l}
\ket{\OM} = \ket{E,\la}\es
a_i^+ \ket{\OM} = \ket{E,\la+\dfrac{1}{2};i}\es
a_i^+ a_j^+ \ket{\OM} = \ket{E,\la+1;[ij]}\es
\cdots\es
a_{i_1}^+ a_{i_2}^+ \cdots a_{i_N}^+  \ket{\OM} = 
\ket{E,\la+\dfrac{N}{2};[i_1\cdots i_N]}
\ea\eqn{ext-spectrum}
If one wants to specialize to renormalizable theories, 
on has to restrict the helicity values to $|\la|\le1$
(supersymmetric gauge theories). One sees, by 
putting $\la=-1$ in the equations above, that the number of supercharges
is restricted to $N\le4$. If one allows for helicities
up to 2 (supergravity), one gets the restriction $N\le8$.

\newpage
%*********************************************************** 
\section{Field Supermultiplets. Models}\label{superm.-de-champs}
\biblio{gates-book,bagger-book,srivatsana,west-book,ps-book,susy96}
\oilbib
The main object of this Section, after an introduction to the
chiral supermultiplet and of the model of Wess-Zumino exhibiting the
self-interaction of such a supermultiplet, will be the study of
supersymmetric gauge theories, in the Wess-Zumino gauge, and of their
renormalizability. 

Let us begin with a brief historical review. Since their 
discovery~\cite{sym-theories}, most of the efforts towards the
renormalization of the super-Yang-Mills (SYM) theories, for many years,  
have been using the formalism of superspace and
superfields, where the supersymmetry is realized linearly. 
Despite of earlier attempts~\cite{bm}, it is only 
recently~\cite{white2,white1,maggiore,mpw1,mpw2} 
that their renormalization in the Wess-Zumino gauge 
has been successfully achieved\footnote{Let us mention however the
pioneering works of Ref.~\cite{sohnius-west} and of 
\cite{mandelstam-brink}, where strong arguments showing the 
ultraviolet finiteness
of the $N=4$ SYM theory have been presented for the first time.}. 
The latter formalism implies a
supersymmetry which is realized nonlinearly and whose commutation 
relations with the gauge transformations are nontrivial.
But it has the advantage of 
involving a minimum number of unphysical gauge of freedom, on the one
hand, and of being more suitable for applications to theories with
extended ($N>1$) supersymmetry]~\cite{white2,maggiore}, on the other
hand. 

%************************************************************
\subsection{The Chiral Supermultiplet}\label{mult.-chiral}
As it is known since a long time, in field theory 
the number of independent fields 
generally exceeds the number of physical degrees of freedom. (This is
particularly striking in gauge theories, where all sorts of ghost fields
appear.) 
We shall encounter this already in the simplest supersymmetric model in
4-dimensional space-time, namely the Wess-Zumino model with a ``chiral
supermultiplet''.

The chiral supermultiplet consists of one complex scalar field
$A(x)$, one Weyl spinor field $\p_\a(x)$, $\a=1,2$ and a second complex
scalar field $F(x)$. 
The infinitesimal supersymmetry transformations of these fields read
(with infinitesimal anticommuting parameters $\e_\a$, $\eb_\ad$)
\[
\d_{\rm SUSY}\vf = i\lc \e^\a Q_\a 
     + \eb_\ad\QB^\ad,\, \vf \rc = 
  \lp\e^\a\d_\a + \eb_\ad \db^\ad\rp\vf\ ,
\quad \vf = A,\p,F\ ,
\]
with
\eq\ba{ll}
\d_\a A = \p_\a       \ ,          &\db_\ad A = 0\ , \es
\d_\a\p^\b = 2\d^\b_\a F \ ,\qquad
        &\db_\ad\p_\a = 2i\s^\m_{\a\ad}\pa_\m A\ , \es
\d_\a F = 0   \ ,
        &\db_\ad F = i\pa_\m\p^\a \s^\m_{\a\ad}\ . 
\ea\eqn{comp-chir-transf}
and similarly for the complex conjugate components $\AB$, $\psb$ and
$\bar F$. One easily checks that the Wess-Zumino algebra \equ{alg-wz} 
holds:
\eq
\{ \d_\a,\db_\ad \} = 2i\s^\mu_{\a\ad}\pa_\m\ ,\quad
\{ \d_\a,\d_\b \} = \{ \db_\ad,\db_\bd \} = 0\ ,
\eqn{alg-delta}
the infinitesimal translations $x^\m$ $\rightarrow$ $x^\m+\xi^\m$
 being represented by
\[
\d_{\rm transl}\vf = i\lc \xi^\m P_\m,\vf \rc\ ,\quad\mbox{with }
i\lc P_\m,\vf \rc = \pa_\m\vf\ .
\]
\remarks
\item[1.] The space-time integral of the last component 
of a chiral supermultiplet, i.e. of its $F$-component, 
is invariant under supersymmetry:
\eq
\d_{\rm SUSY} \xint F(x) = 0\ .
\eqn{inv-F-comp}
\item[2.] The condition 
\eq
\db_\ad A = 0
\eqn{chiral-cond}
defines the {\it chiral} supermultiplet. The remaining transformation rules 
follow uniquely
from the algebra. Indeed, let us first define the fields $\p_\a$ and $F_{\a\b}$
by
\[
\d_\a A= \p_\a\ ,\quad \d_\a \p_\b = F_{\a\b} = -2\e_{\a\b}F + G_{(\a\b)}\ ,
\]
where we have separated the bi-spinor $F_{\a\b}$ into its antisymmetric and
symmetric parts. It follows immediately from 
the anticommutativity of the operators $\d_a$, that any product of 
three or more of them is identically vanishing. Hence $\d_\a F_{\a\b}=0$, or:
\[
\d_a F =0\ ,\quad \d_a G_{(\a\b)} = 0 \ .
\]
On the other hand, the anticommutation rule of $\db_\ad$ with $\d_\a$ implies
\[\ba{l}
\db_\ad \p_\a = \db_\ad\d_\a A = \lac \db_\ad,\d_\a \rac A = 
  2i\s^\m_{\a\ad}\pa_\m A \ ,\es
\db_\ad F_{\a\b} = \db_\ad\d_\a \p_\b = 
  \lac\db_\ad,\d_\a\rac \p_\b - \d_a\db_\ad \p_\b  \es
\phantom{\db_\ad F_{\a\b}}
  = 2i\lp \s^\m_{\a\ad}\pa_\m\p_\b - \s^\m_{\b\ad}\pa_\m\p_\a \rp \ ,
\ea\]
The right-hand side of the second equation being antisymmetric in the indices
$\a$ and $\b$, we can write
\[
\db_\ad G_{(\a\b)} =0\quad\mbox{and}\quad 
 \db_\ad F =  i\s^\m_{\a\ad}\pa_\m \p^\a\ ,
\]
where we have used that $F=\frac{1}{4}\e^{\a\b}F_{\a\b}$. 
We have thus obtained the transformation rules \equ{comp-chir-transf} 
for the fields $A$, $\p$ and $F$. Moreover, the field $G_{(\a\b)}$ has 
turned out to be
invariant under both $\d_\a$ and  $\db_\ad$, which implies,
through the algebra, that it is a constant:
being uninterested in constant fields, we may as well set it to zero.
\skramer
The structure of the chiral supermultiplet and
of its complex conjugate, the 
antichiral supermultiplet $\{\AB,\psb_\ad,\bar F\}$ defined by 
the condition conjugate to \equ{chiral-cond}: $\d_\a\AB=0$, 
are illustrated in 
Tables~\ref{tab-chiral} and \ref{tab-antichiral}.
%++++++++++++++++++++++++++++++++++++++++++++++++++++++++++++
\begin{table}[hbt]
\centering
{\large
\begin{tabular}{ccccccc}
%\hline
        &           &   $Q$           &            &   $Q$           &   \\
        & $A$       &$\longrightarrow$&$\p_\a$     &$\longrightarrow$&$F$\\
$\bar Q$&$\downarrow$&                &            &                 &   \\
        & $0$       &                 &            &                 &
\end{tabular}
}
\caption{Chiral supermultiplet. The horizontal and vertical
arrows indicate the action of the supercharges $Q_\alpha$ and 
$\bar Q_{\dot\alpha}$,
respectively (see \equ{comp-chir-transf}).
Only the transformations leading to new fields, and not to 
derivatives of them, are represented.}
\label{tab-chiral}
\end{table}
%+++++++++++++++++++++++++++++++++++++++++++++++++++++++++++++++

%++++++++++++++++++++++++++++++++++++++++++++++++++++++++++++
\begin{table}[hbt]
\centering
{\large
\begin{tabular}{ccccccc}
%\hline
        &                &   $Q$            &    \\
        & $\bar A$       &$\longrightarrow$ & $0$\\
$\bar Q$&$\downarrow$    &                  &    \\
        & $\psb_\ad$     &                  &    \\
$\bar Q$&$\downarrow$&                      &    \\
        & $\bar F$       &                  &    
\end{tabular}
}
\caption{Antichiral supermultiplet. The horizontal and vertical
arrows indicate the action of the supercharges $Q_\alpha$ and 
$\bar Q_{\dot\alpha}$,
respectively (see \equ{comp-chir-transf}).
Only the transformations leading to new fields, and not to 
derivatives of them, are represented.}
\label{tab-antichiral}
\end{table}
%+++++++++++++++++++++++++++++++++++++++++++++++++++++++++++++++

The $R$-transformation belonging to the Wess-Zumino superalgebra 
\equ{alg-wz}--\equ{com-r-susy} acts on the chiral supermultiplet and 
on its conjugate as
\[
i\lc \eta R,\vf \rc = \d_R \vf \ ,\quad \vf=A,\p,F 
                                    \mbox{ and conjugates} \ ,
\]
where  $\eta$ is an infinitesimal parameter, and
\eq\ba{ll}
\d_R A = i n A\ ,\qquad& \d_R\AB= -in\AB\ ,\es
\d_R \p_\a = i (n+1)\p_\a\ ,\qquad& \d_R\psb_\ad= -i(n+1)\psb_\ad\ ,\es
\d_R F = i (n+2) F\ ,\qquad& \d_R\bar F= -i(n+2)\bar F\ .
\ea\eqn{chiral-R-transf}
The real number $n$ appearing in the $R$-transformations is called 
the $R$-{\it weight} of this chiral supermultiplet.

%******************************************************************
\subsubsection*{Construction of invariant actions:}
The construction of supersymmetry invariants proceeds from the 
observation that the last component of a supermultiplet always 
transforms under supersymmetry \equ{comp-chir-transf}
as a total derivative, which implies 
that its space-time integral is invariant. We have seen this explicitely
for the chiral multiplet (see Equ.~\equ{inv-F-comp}). 
This remains true for the last component -- the ``$D$-component'' --
 of the most general supermultiplet, called
the {\it vector supermultiplet},
\eq
\f = \lac C,\,\chi_\a,\,\bar\chi_\ad,\,M,\,\bar M,\,v_\m,\,
          \la_\a,\,\lb_\ad,\,D \rac\ ,
\eqn{vect-s-mult}
with the transformation rules
\eq\ba{ll}
\d_\a C = \chi   \ ,\quad \d_\a \chi^\b = \d_\a^\b M   \ ,\quad 
\d_\a\bar\chi_\ad = \smuaad (v_\m+i\pa_\m C) \ ,\es 
\d_\a M = 0  \ ,\quad
\d_\a \bar M = \la_\a-i(\s^\m\pa_\m\bar\chi)_\a    \ ,\quad
\d_\a v_\m = \half(\s_\m\bar\la)_\a 
           - \dfrac{i}{2}(\s^\n\sib_\m\pa_\n\chi)_\a \ ,\es 
\d_\a\la^\b = \d_\a^\b D + i(\s^\n\sib^\m)_\a{}^\b\pa_\n v_\m
   \ ,\quad \d_\a\bar\la_\ad = i\smuaad\pa_\m M  \ ,\quad
\d_\a D = -i(\s^\m\pa_\m\bar\la)_\a \ .
\ea\eqn{vector-susy-transf}
(We consider here the {\it real} vector supermultiplet).
It is not difficult to check that these transformation rules 
are the most general ones compatible with the Wess-Zumino 
algebra, the
reality condition being taken into account. The latter condition allows to 
compute the transformations under $\QB_\ad$ as the complex conjugate 
of the transformations above. The result is illustrated in 
Table~\ref{tab-vector}.
%++++++++++++++++++++++++++++++++++++++++++++++++++++++++++++
\begin{table}[hbt]
\centering
{\large
\begin{tabular}{ccccccc}
%\hline
        &           &   $Q$           &            &   $Q$           
&   \\
        & $C$       &$\longrightarrow$&$\chi_\a$   &$\longrightarrow$
&$M$\\
$\bar Q$&$\downarrow$&                &$\downarrow$&                 
&$\downarrow$ \\
        & $\bar\chi_\ad$&$\longrightarrow$&$v_\m$  &$\longrightarrow$
&$\bar\la_\ad$\\
$\bar Q$&$\downarrow$&                &$\downarrow$&                 
&$\downarrow$ \\
        & $\bar M$   &$\longrightarrow$&$\la_\a$  &$\longrightarrow$
&$D$
\end{tabular}
}
%\caption{ESSAI}
\caption{Real vector supermultiplet. The horizontal and vertical
arrows indicate the action of the supercharges $Q_\alpha$ and 
$\bar Q_{\dot\alpha}$,
respectively (see \equ{vector-susy-transf}).
Only the transformations leading to new fields, and not to 
derivatives of them, are represented.}
\label{tab-vector}
\end{table}
%+++++++++++++++++++++++++++++++++++++++++++++++++++++++++++++++
\remark
One easily checks that any supermultiplet whose first component 
$A$, $\AB$ or $C$ is zero, 
is identically vanishing. This implies that any supermultiplet is 
uniquely defined by its first conponent.
\kramer

In order to write invariant actions one has to consider supermultiplets 
which are themselves products of elementary
supermultiplets, then take their last component $F$, $\bar F$ or $D$,
and integrate. The supermultiplet composition rules for chiral 
supermultiplets are listed here:
\begin{enumerate}
\item[i)] The product $A_1A_2$ of two chiral supermultiplets 
$A_1$ and $A_2$ is
the -- unique -- chiral sup\-er\-mul\-tip\-let\footnote{We often denote a 
supermultiplet with the name of its first component.}
\eq\ba{l}
\lac A_1A_2,\, A_1A_2|_\p,\, A_1A_2|_F\rac  \es\quad
= \lac A_1A_2,\; A_1\p_{2\a}+\p_{1\a}A_2,\; A_1F_2+F_1A_2
  -\half\p_1^a\p_{2\a} \rac \ .
\ea\eqn{chir-chir-prod} 
\item[ii)] The product $A\AB$ of a chiral supermultiplet $A$ by its 
conjugate $\AB$ is the 
-- unique -- real vector supermultiplet (we only write out 
explicitely its first and last component)
\eq\ba{l}
\lac A\AB,\, \cdots,\, A\AB|_D\rac = \es\quad
\lac A\AB,\; \cdots, \right.\es\qquad\left.
 2\pa^\mu A\pa_\mu \AB - A\pa^2\AB - \pa^2A\AB
   + i\p^\a\s^\m_{\a\ad}\pa_\m\psb^\ad
  - i\pa_\m\p^\a\s^\m_{\a\ad}\psb^\ad + 4F\FB    \rac \ .
\ea\eqn{chir-antichir-prod}
%\item[iii)] It is also possible\footnote{We shall however not make use
%of this possibility in these lectures.}
% to create new supermultiplets by the operation
%{\it covariant derivative}\footnote{This operation corresponds 
%to the covariant derivative of the superspace formalism.}: 
%the covariant derivative of a supermultiplet $\f$
% is the -- unique -- supermultiplet $D_\b\f$ or $\DB_\bd\f$ 
%whose first component is the supersymmetry transform $\d_\b$
%or $\db_\bd$   of the 
%first component of $\f$. For example, starting from the supermultiplets
%$A$ \equ{comp-chir-transf} and $C$ \equ{vector-susy-transf}, we have
%\[\ba{ll}
%D_\b C = \lac \chi_\b,\, \cdots,\,\rac\ ,\quad&
%   \DB_\bd C = \lac \bar\chi_\bd,\, \cdots,\,\rac\ ,\es
%D_\b A = \lac \p_\b,\, \cdots,\,\rac\ ,\quad&
%   \DB_\bd A = \lac 0,\, \cdots,\,\rac = 0\ .
%\ea\]
%The covariant derivatives obey the algebra
%\eq
%\lac D_\a,\DB_\ad \rac = 2i\s^\mu_{\a\ad}\pa_\m\ ,\quad
%\lac D_\a,D_\b \rac = \lac \DB_\ad,\DB_\bd \rac = 0\ ,
%\eqn{alg-cov-der}
%which follows from the supersymmetry algebra \equ{alg-delta}.
\end{enumerate}
%**************************************************************
\subsubsection*{The Wess-Zumino model:}
Let us consider now, as an illustration, the theory of one chiral 
supermultiplet $A$. 
Assigning canonical dimensions 1 and 3/2 to the fields $A$ and $\p$, 
respectively -- then  $F$ has dimension 2  --
the most general supersymmetric action 
which is power-counting renormalizable, i.e. 
of dimension bounded by
four\footnote{Our units are chosen such that $c=\h=1$. Dimensions are
counted in mass units. The ``dimension'' of a space-time integral such
as the action is by definition the dimension of its integrand -- here: the
Lagrangian density.}
reads
\eq\ba{l}
\S(A,\AB) = \xint \lp \dfrac{1}{4} \left.A\AB\right|_D 
 - \dfrac{m}{2}\lp \left.A^2\right|_F + \left.\AB^2\right|_\FB \rp
 - \dfrac{\la}{12}\lp \left.A^3\right|_F + \left.\AB^3\right|_\FB \rp \rp\es
\phantom{\S(A,\AB)} = 
\xint \lp \pa^\m A \pa_\m\AB + \dfrac{i}{2}\p\s^\m\pa_\m\psb 
+ F\bar F     \right.\es\phantom{S(A,\AB)= }\left.
    -\dfrac{m}{4}(4AF-\p^2 + \mbox{ conj.})
  -\dfrac{\la}{8}(2A^2F-A\p^2+\mbox{ conj.}) \rp
\ea\eqn{baby-action-comp}
This action is moreover invariant under the $R$-transformations
\equ{chiral-R-transf}, up to the mass terms, if we assign the 
$R$-weight $n=-2/3$ to the chiral multiplet $A$.

The field equations are\footnote{{\bf Notation:} $(\sla\pa\psb)_\a
= \s^\m_{\a\ad}\pa_\m\psb^\ad$.}
\eq\ba{l}
\dfud{\S}{A} = -\pa^2 \AB-mF -\dfrac{\la}{2}AF = 0 \ ,\es
\dfud{\S}{\p} = \dfrac{i}{2}\sla\pa\psb + \dfrac{m}{2}\p +
 \dfrac{\la}{4} A\p = 0\ ,\es
\dfud{\S}{F} = \FB - m A - \dfrac{\la}{4} A^2 = 0 \ .
\ea\eqn{baby-eq-motion}
It is worth noticing that $F$ and $\FB$ are auxiliary fields, 
i.e. that their equations of motion are purely algebraic.
It follows that they can be eliminated and thus effectively do not 
represent independent degrees of freedom. 
Substituting the expression given by \equ{baby-eq-motion} 
for $F$ and its conjugate
in the action \equ{baby-action-comp}, 
we get the equivalent action  
\eq\ba{l}
\S(A,\AB)=\es\ \xint\lp \dfrac{i}{2}\p\s^\m\pa_\m\psb + 
\pa^\m A\pa_\m\bar A 
 +\dfrac{m}{4}\p^2+\dfrac{m}{4}{\psb}^2+
\dfrac{\la}{8}A\p^2+\dfrac{\la}{8}{\bar A}\psb^2 -
V(A,\bar A)\rp \ ,
\ea\eqn{baby-action-elim}
with a potential given by
\eq
V(A,\bar A) = F(\bar A)\bar F(A) 
  = \left\vert mA+\dfrac{\la}{4}A^2\right\vert^2   \ ,
\eqn{baby-pot}
which turns out to be positive. This positivity is a pecularity of
supersymmetric theories.
\remarks
\item[1.] The new action \equ{baby-action-elim} is invariant under the 
transformation laws \equ{comp-chir-transf} where $F$ has been eliminated
 as well. The algebra \equ{alg-delta} is however modified: 
the anticommutator
of two supersymmetries is no more  a simple translation, but there is 
an additional term involving a field equation. This happens for the 
anticommutator as applied to the spinor component only:
\eq
\lac \d_\a,\db_\ad \rac\p^\b = 2i\s^\m_{\a\ad}\pa_\m\p^\b
  + 4\d^\b_\a \dfud{\S}{\psb^\ad}\ .
\eqn{anticom-psi}
\item[2.] This nonclosure, and the fact that the transformations 
are now nonlinear, could 
 complicate the renormalization procedure: the fields transform now 
into composite
 fields, which may be the source of
 new ultraviolet divergences. It is thus useful, in certain cases,
 to keep oneself from eliminating the auxiliary field $F$.
\skramer
%************************************************************
\subsection{The Vector Supermultiplet. Gauge 
Invariance}\label{inv-de-jauge}
The  transformation rules of the ``real vector supermultiplet''
have been given 
by \equ{vector-susy-transf} and illustrated in Table~\ref{tab-vector}.
This multiplet containing a vector field, $v_\m$, is a good candidate
for a supersymmetric generalization of the concept of
gauge vector field. There is however another possibility of
implementing supersymmetry in gauge theories, which involves a shorter
supermultipet, the  Wess-Zumino  gauge supermultiplet, in short: the  
{\it WZ-multiplet}. It consists of a vector field $A_\m$, of a Weyl
spinor field $\la_\a$ and of an auxiliary pseudoscalar field $D$.
Such a supermultiplet only makes sense, as we shall see, in the context
of a gauge theory. 

Let us write the 
supersymmetry transformation laws in the case we are dealing with
a general nonabelian gauge theory. 
The generators $X_a$ of the gauge group $G$
obey the Lie algebra 
\eq
\lc X_a,X_b \rc = i f_{ab}{}^c X_c\ ,
\eqn{lie-alg}
and the fields of the WZ-multiplet take an index $a$,
the infinitesimal gauge transformations being
\eq
\d_{\rm gauge} A_\m^a = \nabla_\m\om^a = 
 \pa_\m\om^a + f_{bc}{}^a A_\m^b\om^c\ ,
\d_{\rm gauge} \vf^a = f_{bc}{}^a \vf^b \om^c\ ,\quad
 \vf = \la,\lb,D\ ,
\eqn{gauge-trf}
where $\om^a(x)$ is the infinitesimal local parameter.
Then the supersymmetry transformations read
\eq\ba{l}
\d_\a A^a_\m = \dfrac{1}{4} \s_{\m\a\ad}\lb^{a\ad}\ ,\es
\d_\a \la^{a\b} = \d_\a^\b D^a + 2 \s^{\m\n}{}_\a{}^\b F^a_{\m\n}\ ,\es
\d_\a \lb^a_\ad = 0\ ,\es
\d_\a D^a = -i \s^\m_{\a\ad}\nabla_\m\lb^{a\ad}\ ,
\ea\eqn{WZ-transf}
with the Yang-Mills field strength and the covariant derivative defined
by 
\eq\ba{l}
F^a_{\m\n}=\pa_\m A^a_\n-\pa_\n A^a_\m + f_{bc}{}^a A_\m^b A_\n^c\ ,\es
\nabla_\m\vf^a = \pa_\m \vf + f_{bc}{}^a A_\m^b \vf^c\ ,
   \quad \vf = \la, \lb, D\ .
\ea\eqn{ym-field-cov-der}
These transformation rules are illustrated in Table~\ref{tab-WZ}.
%++++++++++++++++++++++++++++++++++++++++++++++++++++++++++++
\begin{table}[hbt]
\centering
{\large
\begin{tabular}{cccc}
%\hline
        &            &   $Q$           &   \\
        &$A_\m$      &$\longrightarrow$&$\lb_\ad$\\
$\bar Q$&$\downarrow$&                 &$\downarrow$ \\
        &$\la_\a$    &$\longrightarrow$&$D$
\end{tabular}
}
\caption{Wess-Zumino vector supermultiplet. The horizontal and vertical
arrows indicate the action of the supercharges $Q_\alpha$ and 
$\bar Q_{\dot\alpha}$,
respectively (see \equ{WZ-transf}).
Only the transformations leading to new fields -- and not to 
derivatives of them  -- are represented.}
\label{tab-WZ}
\end{table}
%+++++++++++++++++++++++++++++++++++++++++++++++++++++++++++++++
The two main differences between the latter transformation rules and the
rules \equ{vector-susy-transf} are that they are nonlinear in the
fields, on the one hand, and that their algebra is not exactly of the
Wess-Zumino type, as in \equ{alg-delta}, on the other hand. 
Concerning the second point, one checks indeed that these
transformations obey the following superalgebra, which we 
write down here for every component of the WZ-multiplet:
\eq\ba{l}
\lac \d_\a,\db_\ad \rac A_\n^a = 2i\s^\m_{\a\ad}F^a_{\m\n}
 = 2i\s^\m_{\a\ad} \pa_\m A_\n^a-i\nabla_\n \om^a_{\a\ad}\ ,\es
\lac \d_\a,\db_\ad \rac \vf^a = 2i\s^\m_{\a\ad}\nabla_\m\vf^a
 = 2i\s^\m_{\a\ad} \pa_\m \vf^a + f_{bc}{}^a \om^b_{\a\ad}\vf^c\ ,
\quad \vf=\la,\lb,D\ ,
\ea\eqn{s-alg-WZ-gauge}
with
\eq
\om^a_{\a\ad} = 2\s^\m_{\a\ad}A^a_\m\ .
\eqn{field-dep-param}
One recognizes in the right-hand sides the superposition of a
translation and of a gauge transformation \equ{gauge-trf},
but with a {\it field dependent
parameter} $\om^a_{\a\ad}$ as given by \equ{field-dep-param}. 
\remarks
\item[1.] 
The algebra generated by these supersymmetry transformations is not
closed. Indeed, if we compute the commutator
of the right-hand sides of \equ{s-alg-WZ-gauge} with supersymmetry
transformations, we obtain gauge transformations whose parameter
is the supersymmetry transform of \equ{field-dep-param}, and so on,
repeating the operation. This finally leads to an infinite chain of new
gauge transformations, which sets an apparently insolvable problem when
coming to the quantum corrections, each new gauge transformation
involving a new composite field which ought to be 
renormalized~\cite{bm}.
It is however clear that, at the classical level, 
when applied to gauge invariant quantities,
this algebra reduces to the usual Wess-Zumino \equ{alg-delta} one.
A solution of this puzzle, applicable to the quantum theory, will be
presented in Subsection~\ref{alg.-de-BRS}.
\item[2.]
It is possible to work with the full vector supermultiplet 
\equ{vect-s-mult}, with the advantage that its supersymmetric 
transformations \equ{vector-susy-transf} are linear, 
which allows the use of the 
formalism of superfields and 
superspace~\cite{gates-book,ps-book,susy96}. 
However, this multiplet contains supplementary fields 
$C$, $\chi_\a$ and $M$, which represent unphysical gauge degrees 
of freedom associated to a``supergauge invariance''. 
Two options are then possible:
\begin{enumerate}
\item[a)] Either work with all these variables, taking profit of the
power of superspace formalism. But this gain is less obvious if
one wants to treat theories with supersymmetry breaking.
\item[b)] Or eliminate the spurious variables by choosing a special 
supergauge 
called the {\it Wess-Zumino gauge}. The resulting version of the theory 
is precisely the one we are considering in these lectures.
\end{enumerate}
\skramer
%************************************************************
\subsubsection*{Matter fields:}
In the class of theories we are going
to consider, matter is described by
chiral supermultiplets $\lac A^i,\p^i,F^i\rac$, with gauge transformations
\eq
\d_{\rm gauge} \vf^i(x) = -iT_{aj}^i \om^a(x)\vf^j(x)\ ,\quad
  \vf^i= A^i,\p^i,F^i\ ,
\eqn{mat-g-transf}
corresponding to a certain unitary representation of the gauge group,
the matrices $T_a$ obeying the commutation rules \equ{lie-alg}.
The supersymmetry transformations are the ones we have given
previously, \equ{comp-chir-transf}, 
but with derivatives replaced by gauge covariant derivatives 
\eq
\nabla_\m \vf^i = \pa_\m \vf^i + i T_{aj}^i A_\m^a \vf^j\ .
\eqn{mat-cov-der}
They read:
\eq\ba{ll}
\d_\a A^i = \p^i_\a       \ ,          &\db_\ad A^i = 0\ , \es
\d_\a\p^{i\b} = 2\d^\b_\a F \ ,\qquad
        &\db_\ad\p^i_\a = 2i\s^\m_{\a\ad}\nabla_\m A^i\ , \es
\d_\a F^i = 0   \ ,
        &i\db_\ad F^i = i\nabla_\m \p^{i\a} \s^\m_{\a\ad}\ . 
\ea\eqn{mat-susy-transf}
As for the WZ-multiplet, the algebra closes on the translations,
accompanied by field dependent gauge transformations:
\eq 
\lac \d_\a,\db_\ad \rac \vf^i = 2i\s^\m_{\a\ad}\nabla_\m\vf^i
 = 2i\s^\m_{\a\ad} \pa_\m \vf^i + i T_{aj}^i \om^a_{\a\ad}\vf^j\ ,
\quad \vf^i=A^i,\p^i,F^i\ ,
\eqn{mat-s-alg-WZ-gauge}
with $\om^a_{\a\ad}$ given by \equ{field-dep-param}.
%****************************************************
\subsubsection*{Super-Yang-Mills action:}
The most general supersymmetric 
gauge invariant action of dimension bounded by 4,
involving the WZ-multiplet and
the matter chiral supermultiplets, with supersymmetry 
transformations as given by \equ{WZ-transf} and \equ{mat-susy-transf},
reads
\eq\ba{l}
S_{\rm SYM} = \xint \LP 
    \displaystyle{\dfrac{1}{g^2}}\lp
    -\dfrac{1}{4}F^a_{\m\n}F_a^{\m\n}
    +i\la^{a\a}\smab\nabla_\m\lb^{a\bd} + \half D^aD_a\rp \es
\phantom{S_{\rm SYM} =}
+\nabla_\m A^i \nabla_\m\AB_i 
 +\dfrac{i}{2}\p^{i\a}\smab(\nabla_\m\psb^\bd)_i + F^i\FB_i \es
\phantom{S_{\rm SYM} =}
+\lp - \half m_{(ij)} (2A^iF^j - \half \p^i\p^j)
  -\dfrac{1}{4} (\la_{(ijk)}(A^iA^j F^k -\half\p^{i\a} \p^j_\a A^k)
      +\mbox{ conj.} \rp  \es
\phantom{S_{\rm SYM} =}
+\lp -i\psb_{i\bd}T_a^i{}_j A^j\lb^{a\bd } + \mbox{ conj.} \rp \es
\phantom{S_{\rm SYM} =}
+\dfrac{1}{4}\AB_i T_a^i{}_j A^j D^a  \RP\ ,
\ea\eqn{sym-action}
The mass matrix $m_{(ij)}$ and the Yukawa coupling constants
$\la_{(ijk)}$ are invariant symmetric tensors of the gauge group.

This action is also invariant,
up to the mass terms, under the $R$-trans\-for\-mat\-ions, given by 
\equ{chiral-R-transf} with the weight $n=-2/3$ for the chiral multiplets, 
and by 
\eq
\d_R A_\m=0 \ ,\quad \d_R\la_\a=-\la_\a\ ,
                     \quad \d_R\lb_\ad=\lb_\ad\ ,
\eqn{vect-R-transf}
for the WZ-multiplet. The generator $R$ still obeys the commutation
rules \equ{com-r-susy} with the supersymmetry generators, and commutes
with the translation generator.
\remark The supersymmetric extensions of the Standard Model of
elementary interactions are given by actions of the type 
\equ{sym-action}, with  specific gauge
group and representations. This is in  particular the case of the ``Minimal
Supersymmetric Standard Model'' and of the Supersymmetric 
Grand Unified Theories~\cite{super-reviews}.
\kramer
As for the model of subsection~\ref{mult.-chiral}, we can eliminate 
the auxiliary fields $F$, $\FB$ and $D$ by substituting 
their equations of  motion:
\eq\ba{l}
\FB^i(A) = m_{ij} A^j + \dfrac{1}{4}\la_{ijk} A^jA^k\ ,\es
D_a(A,\AB)  = -\dfrac{1}{4} \AB_iT_{aj}^i A^j\ ,
\ea\eqn{aux-fields-sym}
in the action \equ{sym-action} and in the supersymmetry transformation 
laws \equ{vector-susy-transf} and \equ{mat-susy-transf}. Again, 
the action contains a potential term for the scalar fields which is
positive:
\eq
V(A,\AB) = \half D^a(A,\AB)D_a(A,\AB) + F^i(\AB)\FB_i(A) \ .
\eqn{sym-pot}
Also, the superalgebra does close only modulo equations of motions. 
Explicitly, the anticommutator $\{\d_\a,\db_\ad\}$ applied 
on the spinor fields yields (c.f. \equ{anticom-psi})
\eq\ba{l}
\lac \d_\a,\db_\ad \rac\p^\b = 2i\s^\m_{\a\ad}\nabla_\m\p^\b
  + 4\d^\b_\a \dfud{\S_{\rm SYM}}{\psb^\ad}\ ,\es
\lac \d_\a,\db_\ad \rac\la^\b = 2i\s^\m_{\a\ad}\nabla_\m\la^\b
  - 2\d^\b_\a \dfud{\S_{\rm SYM}}{\lb^\ad}\ .
\ea\eqn{anticom-psi-lambda}

%*****************************************************************
\section{Renormalization of SYM Theories in the \hfill\break 
Wess-Zumino Gauge}\label{renormalisation}
The problem to be solved is to show the renormalizability of the
super-Yang-Mills (SYM) 
theory, i.e. to show that it is possible to 
 construct a perturbative quantum version of it, 
preserving both gauge invariance and supersymmetry. Since we don't 
dispose of any  
regularization procedure preserving both invariances together,
it is convenient to use a general, regularization independent
renormalization scheme. Such a scheme, 
called the ``algebraic renormalization''~\cite{p-sor-book},
has  been exposed by Silvio Sorella in his lectures at the present 
School~\cite{sorella-vitoria}. I shall only summarize 
the procedure as I want to apply it 
here, only emphasizing the points which are 
specific to the supersymmetric gauge models. 

In order to simplify at
most the exposition, I shall consider the pure SYM case, without matter
fields, the auxiliary field $D_a$ eliminated -- its equation of motion
reducing  now to the trivial equation $D_a=0$.

%******************************************************************
\subsection{BRS Algebra. Gauge Fixing}\label{alg.-de-BRS}
The starting point is the gauge and supersymmetry invariant action
\equ{sym-action}, without the matter chiral supermultiplets 
$A^i$, and with
the auxiliary fields $D_a$ eliminated:
\eq
S_{\rm SYM} = \xint \dfrac{1}{g^2} \lp
    -\dfrac{1}{4}F^a_{\m\n}F_a^{\m\n}
    +i\la^{a\a}\smab\nabla_\m\lb^{a\bd} \rp 
\eqn{pure-sym-action}
Being masssless, this action is exactly invariant under the 
$R$-transformations \equ{vect-R-transf}.
\remark
The action \equ{pure-sym-action} looks like that of an ordinary gauge
theory, the ``gaugino'' field $\la$ playing the role of a matter field
minimally coupled to the gauge field.
The only specificity, which makes the theory supersymmetric, is that 
$\la$ belongs to the adjoint representation of the gauge group.
\kramer
We know that the superalgebra of the symmetries of the theory is
not closed (c.f. the first of the remarks
following Equ.\equ{gauge-trf}). In
particular, we have seen that the anticommutator of two supersymmetry 
transformations is a translation accompanied by a field dependent 
gauge transformation (and a by a field equation since we have eliminated
the auxiliary fields). And we have observed that this leads
-- apparently -- 
to an infinite algebra involving an infinity of new generators.

The way to solve this puzzle has been found by White~\cite{white1,white2} 
(See
also~\cite{maggiore,mpw1,mpw2}). It consists in putting together all 
the generators of the algebra into one single BRST operator. Let us
recall that the usual BRS symmetry of gauge theories is obtained by
replacing in the gauge transformations the local infinitesimal
parameters $\om^a(x)$ by {\it anticommuting} fields $c^a(x)$, the
Faddeev-Popov ghost fields, and by defining the transformation laws
of these ghost fields in such a way that the transformation be
nilpotent. Calling $s$ the {\it BRS operator} obtained in this way, we
can write the result as
\eq\ba{l}
s A_\m^a = \nabla_\m c^a\ ,\quad  s\la^a = f_{bc}{}^a \la^b c^c\ ,\es
s c^a = -\half f_{bc}{}^a c^b c^c\ ,\qquad s^2=0\ ,
\ea\eqn{BRS-gauge-trf}
with the covariant derivative defined as in \equ{ym-field-cov-der}.

Since the superalgebra we have in the present theory 
mixes gauge with
supersymmetry transformations, it is 
naturel\footnote{At my knowledge, \cite{bbbcd} is the first paper
proposing the use of the BRS techniques for dealing with global 
symmetries. Dixon~\cite{dixon} has applied this idea to supersymmetry.}
 to generalize the BRS 
operator above, including into it the infinitesimal supersymmetry
transformations \equ{WZ-transf}, but with the infinitesimal parameters
$\e^a$ becoming {\it commuting} numbers instead of anticommuting ones.
These ``global ghosts''  are constant, of course, supersymmetry 
not being a local invariance.
The translations and the $R$-transformations \equ{vect-R-transf}
belonging to the algebra, we also promote their
infinitesimal parameters $\xi^\m$ to the rank of global ghosts:
they will now be anticommuting numbers. The general rule is to assign to
the (local or global)  ghosts the statistics opposite to the one of the
usual infinitesimal parameters -- as shown in Table~\ref{tab-inf-param}
for the present theory.
%++++++++++++++++++++++++++++++++++++++++++++++++++++++++++++++
\begin{table}[hbt]
\centering
\begin{tabular}{lll}
%\hline
{\bf Symmetry} &{\bf Generators}$\qquad$  &{\bf Infinitesimal parameters} \\
\hline\\
Supersymmetry$\qquad$ &$Q_\a,\,\QB_\ad$  &$\e^\a,\,\eb^\ad$ (odd) \\
Translations   &$P_\m$            &$\xi^\m$ (even) \\
$R$         &$R$               &$\eta$ (even) \\
Gauge          &$X_a$             &$\om^a(x)$ (even) \\
\end{tabular}
\caption{The symmetries of the super-Yang-Mills
theory, the names of their generators and infinitesimal
parameters, and the grading, i.e. the statistics, of the latters.}
\label{tab-inf-param}
\end{table}
%+++++++++++++++++++++++++++++++++++++++++++++++++++++++++++++++

As in the case of the pure gauge BRS invariance \equ{BRS-gauge-trf}, we
ought to define the transformation rules for the global as well as for 
for the local ghosts in
order to achieve the nilpotency of the generalized BRS operator.
It turns out however that, due to the superalgebra \equ{alg-wz} 
having field
equations in its right-hand sides because of the elimination of the
auxiliary fields (see \equ{anticom-psi-lambda}), 
the generalized BRS operator will only be nilpotent up
to field equations.
Denoting the generalized BRS operator with the same symbol $s$ as above, 
we have
\eq\ba{l}
s A^{a}_\m =  \nabla_\m c^a
+\e^\a\s_{\m\a\bd}\lb^{a\bd} +\la^{a\a}\s_{\m\a\bd}\eb^\bd
           - i\xi^\n\pa_\n A^{a}_\m                         \es
s \la^{a\a}   = -f^{abc} c^b \la^{c\a}
                -\dfrac{1}{2}\e^\g\s^{\m\n}{}_\g{}^\a F^a_{\m\n}
             -\dfrac{i}{2}g^2\e^\a(\AB_i T_a^i{}_j A_b)
           - i\xi^\m\pa_\m \la^{a\a} -\eta\la^{a\a}          \es
sc^a = -\dfrac{1}{2}f^{abc}c^bc^c -2i\esmab A^a_\m -i\xi^\m\pa_\m c^a \es
s\xi^\m = -2\esmab                                             \es
s\e^\a=-\eta\e^\a                                              \es
s\eta = 0\ ,
\ea\eqn{brs-gener}
with
\eq
s^2= \mbox{ field equations}\ .
\eqn{brs-nilpotency}
The right-hand side of \equ{brs-nilpotency} is effectively
nonzero for $s^2$ applied to the spinor
fields, as a consequence of \equ{anticom-psi-lambda}:
\[
s^2 \la^b = 2\e^\b\eb^\ad \dfud{\S_{\rm SYM}}{\lb_\ad}\ .
\]

%************************************************************
\subsubsection*{Gauge fixing:}
A prerequisite to the construction of a quantum extension of the
classical theory defined by the gauge invariant supersymmetric
action \equ{pure-sym-action} is the fixation of the gauge. In the BRS
framework, this is most conveniently done by introducing a 
{\it Lagrange multiplier} field $B_a(x)$ and a {\it Faddeev-Popov 
antighost} field $\cb_a(x)$, with BRS transformations
\eq
s\cb^a =  B^a -i\xi^\m\pa_\m\cb^a\ ,\quad
sB^a = -2i \esmab\pa_\m\cb^a -i\xi^\m\pa_\m B^a\ .
\eqn{brs-antighost}
The terms in  $\xi$ and $\e$ assure the compatibility with the BRS
transformations previously defined in \equ{brs-gener}, together with the
preservation of the nilpotency.

The gauge is then fixed by adding to the action a suitable BRS invariant 
piece $\S_{\rm gf}$ 
containing the fields $B$ and $\cb$. A particular choice is the {\it Landau
gauge}, defined by  
\eq
\S_{\rm gf} = s \xint  \lp \cb_a\pa^\m A_\m^a \rp
= \xint  \lp B_a\pa^\m A_\m^a 
+ \cb_a \pa^\m \nabla_\m c^a \rp + O(\epsilon,\bar\epsilon)\ .
\eqn{gf-action}
The BRS invariance of \equ{gf-action}
follows from the nilpotency of $s$, which here is exact 
since no spinor field is involved. The Lagrange multiplier term 
induces the gauge fixing constraint $\pa^\m A_\m^a=0$
on the gauge field. 

\noindent The total action
\eq
\S = \S_{\rm SYM}+ \S_{\rm gf}
\eqn{tot-action}
will be the starting point of the perturbative quantization.

%*****************************************************************
\subsection{Renormalization. Anomaly}\label{ren-anomalie}
\biblio{white1,mpw1,mpw2}
\oilbib
The renormalization of a given classical theory -- in our case, the
theory defined in the preceding Subsection --
consists in the perturbative construction of a quantum theory 
which coincides with it in the limit of vanishing Planck 
constant~\cite{p-sor-book,sorella-vitoria}. ``Perturbative'' 
means here an expansion\footnote{The expansion in $\h$ 
is equivalent here to the
expansion in the powers of the gauge coupling constant $g$.} 
in the powers of $\h$, this power corresponding to the number of loops
in the Feynman graphs.

The construction of the quantum theory 
should preserve the BRS invariance of the classical theory. If the task
turns out to be impossible, one speaks of an anomaly. 

The quantum BRS invariance
is expressed by a {\it Slavnov-Taylor identity} which,
in the classical limit, corresponds to the BRS invariance of the 
action \equ{tot-action}. 
At the quantum level, the relevant object is the Green  functional,
i.e. the generating functional of the Green functions\footnote{The
symbol $T$ stands for ``time-ordered product'', the indices $A_k$ stand
for the field type and for the values of the internal and Lorentz
indices and of the space-time cooordinates alltogether. Note that we
are using the same symbol $\vf_{A_k}$ for 
denoting an argument of the vertex
functional -- a smooth function -- and the corresponding quantum field
operator.} 
$\bra{0}T\,\vf_{A_1}\cdots\vf_{A_N}\ket{0}$. An equivalent
object -- more convenient for the purpose of renormalization --
is the {\it vertex functional} $\G(\vf)$, which generates the
vertex functions $\bra{0}T\,\vf_{A_1}\cdots\vf_{A_N}\ket{0}_{\rm 1PI}$, 
i.e. the Green functions amputated of their external
lines and to which only the one-particle irreducible 
(1PI) Feynamn graphs\footnote{A 1PI graph is a Feynman graph with all
its external lines suppressed, and which moreover is
such that it cannot be decomposed in
two disjoint pieces by cutting one single line. Any 
general Green function has
the structure of a tree, the lines representing the full propagators 
(the 2-point Green functions), and the vertices corresponding to the 
vertex functions. The potential ultraviolet singularities caused by 
divergent integrals on the loop momenta are contained in the 1PI graphs
contributing to the vertex functions.} 
are contributing:
\[
\left.\dfrac{\d\G}{\d\vf_{A_1}\cdots\d\vf_{A_N}}\right|_{\vf=0}
= \bra{0}T\,\vf_{A_1}\cdots\vf_{A_N}\ket{0}_{\rm 1PI}\ .
\]
The perturbative expansion of the vertex functional:
\eq
\G(\vf) = \dsum{n\ge0}{} \h^n \G_n(\vf)\ ,
\eqn{exp-gamma}
is such that it coincides with the classical action in the classical
limit\footnote{Which corresponds to the approximation of 
the tree Feynman graphs, i.e. of the graphs without loop}:
\eq
\G_0(\vf) = \S(\vf) \ ,\quad\mbox{or:}\quad \G(\vf)=\S(\vf)+O(\h)\ .
\eqn{vertex-funct}

Let us write, symbolically\footnote{A precise 
formulation~\cite{p-sor-book,sorella-vitoria}  would  need the 
introduction of external fields -- the Batalin-Vilkovisky 
``antifields'' -- coupled to the BRS variations of the various quantum 
fields. It is in this formalism that the definition of
the BRS operator can be extended in such a way 
that it becomes exactly nilpotent:
$s^2=0$.},
the Slavnov-Taylor identity to be proved as
\eq
s \G =0\ .
\eqn{slavnov}
In the classical limit this reduces to the mere BRS
invariance of the action:
\eq
s \S =0\ .
\eqn{class-slavnov}
Let us sketch the inductive proof of \equ{slavnov}.
We assume it to hold up to and including the order $n-1$, for $n\ge1$:
\eq
s \G =O(\h^{n}) \ .
\eqn{slavnov(n-1)}
In order to proceed, let us use the ``quantum action principle'' 
(see e.g.~\cite{p-sor-book}). This theorem
states that a variation of the vertex functional -- such as the one
given here by the BRS operator -- yields a local insertion, i.e.:
\[
s\G = \D\cdot\G = \D + \mbox{loops}\ ,
\]
where $\D$ is the space-time integral of a local field polynomial. In
the second equality we have separated out the contributions of the
trivial tree graphs consisting 
only of the single vertices corresponding to the
field polynomial $\D$, from the loop graph contributions, which are of
higher order in $\h$.
Making explicit the factor in $\h$ following from the 
assumption \equ{slavnov(n-1)}, we can thus write
\[
s\G = \h^n \D + O(\h^{n+1})\ .
\]
The nilpotency of $s$ implies for $\D$ the {\it consistency condition} 
\eq
s\D=0\ .
\eqn{cons-cond}
For the present theory, the most general local field polynomial 
satisfying this condition has
the form
\eq
\D = s\hat\D \ ,
\eqn{hat-delta}
where $\hat\D$ is an integrated local field polynomial of dimension  
4, i.e. of the dimension of the action.
Redefining the action by adding $-\hat\D$ as a counterterm:
\eq 
\S \longrightarrow  \S' =\S - \h^n\hat\D\ , 
\eqn{redef-action}
leads to a modification of the vertex functional:
\[ 
\G \longrightarrow  \G' = \G - \h^n\hat\D + O(\h^{n+1})\ , 
\]
such that the Slavnov-Taylor identity now holds to the next order:
\eq
s\G = O(\h^{n+1})\ .
\eqn{slav-next-ord}  
This ends the inductive proof of the renormalizability of the pure
super-Yang-Mills theory.
\remark The proof given here has been very sketchy. 
Beyond the fact that we have not written the Slavnov-Taylor in a proper
way, using external fields coupled to the BRS variations of the quantum
fields, we have not checked the stability of the gauge fixing,
which turns out to hold unchanged to all orders. 
A complete demonstration may be found in~\cite{mpw1}.
\kramer
%*********************************************************
\subsubsection*{Anomaly:}
The result above holds for the theory without matter 
(described by chiral supermultiplets). In presence of matter, the
right-hand side of \equ{hat-delta} may contain a
supplementary term which is not the BRS variation of an
integrated local field polynomial, and thus cannot be reabsorbed as a
counterterm in the action. This term is the {\it gauge anomaly}, which,
if present, would definitively break BRS invariance. It has the form
\eq
\D = r \xint\lp \e^{[\m\n\rho\s]} d_{(abc)} 
 \lp \pa_\m c^a - 2\la^a\s_\m\eb - 2\e\s_\m\lb^a \rp A_\n^b\pa_\rho A^c_\s 
 + \cdots \rp\ ,
\eqn{anomaly}
where $\e^{[\m\n\rho\s]}$ is the completely antisymmetric Levi-Civita
tensor, and $d_{(abc)}$ a completely symmetric invariant tensor of the
gauge group. The dots represent contributions of higher order in the
number of fields\footnote{See~\cite{mpw1} 
for the derivation and the complete expression of this anomaly.}.

Whereas the form of the anomaly is independent of the details 
of the theory, in particular of its matter content, the coefficient $r$
depends from the latter. Its value at first order in $\h$
indeed is given by the calculation of the 1-loop fermion triangle graph
of the Adler-Bardeen type -- exactly like in usual gauge theories, 
and is proportional to 
$d_{(abc)}\tr(T^aT^bT^c)$, where the matrices $T^a$ are the generators
of the gauge group in the matter field representation. As in the 
usual gauge theories also,  it is possible to make 
the coefficient $r$ vanish at first order by
suitably choosing this representation, and also here
(see~\cite{mpw1}) a
nonrenormalization theorem assures then its vanishing 
to all higher orders.

%*****************************************************************
\section{Broken Supersymmetry}\label{susy-brisee}
As it is well known, nature is not supersymmetric, at least in the
domain of energies accessible to present experiments. 
In particular, no boson-fermion particle doublet with degenerate mass
has been observed as yet.
Therefore
supersymmetry must be broken. One may consider two ways of doing it,
either {\it explicitly}, by adding e.g. nonsupersymmetric mass terms in the
action, or {\it spontaneously}. The difficulty with the latter option is
that it necessarily implies the existence of a 
massless Goldstone particle, which must be a fermion
due to the fermionic nature of the supersymmetry generators. The only
known candidates are the  neutrinos. But low energy theorems -- analogous
to those for the pions in the framework of the chiral effective 
theories~\cite{chiral-eff-th} -- would suppress low energy phenomena
such as $\beta$ decay in an unacceptable way. One is thus left with the
other term of the alternative, i.e. 
with the explicit supersymmetry breakdown 
by mass terms. I shall however also present a model with spontaneous
breakdown of supersymmetry, the O'Raifeartaigh model~\cite{o'raif}
for its own theoretical interest: it shows indeed a mechanism through
which a particle which is massless in the classical theory can acquire a
(calculable) mass through the radiative corrections.

%*****************************************************************
\subsection{Explicit Supersymmetry Breaking}
The lack of supersymmetry is most apparent in the fact that
supermultiplets of particles, if they exist at all, must contain states
of different masses. Indeed, the fact that no supersymmetric partner of
any known particle has yet been observed implies, for these
hypothetical particles, masses of the order of $10^2$ GeV at least. 
A rather obvious way to achieve this goal is to introduce mass terms for 
each of them in the action, thus breaking supersymmetry explicitly. 

Such breakings are soft in the sense that they don't influence the high
energy-momentum behaviour of the theory, leaving in particular 
supersymmetry intact in the asymptotic region. But one would also like 
that the breakings don't hurt a property which singles out
supersymmetric theories: the absence of quadratic 
divergences~\cite{gates-book,ps-book}. 
This property assures the stability of the huge differences
of scales which one encounters in the models of Grand 
Unification~\cite{super-reviews}.

Girardello and Grisaru~\cite{gir-gri} have shown that a
certain class of mass terms -- which they also called ``soft breakings'' --
do break supersymmetry, but preserve the absence of quadratic
divergences. The ``soft breakings'' relevant for the supersymmetric
gauge theories described in Section~\ref{superm.-de-champs}, and which
may be added to the action \equ{sym-action}, are listed in
Table~\ref{tab-soft-br} together with interactions terms sharing the same
properties. 
%++++++++++++++++++++++++++++++++++++++++++++++++++++++++++++++
\begin{table}[hbt]
\centering
\begin{tabular}{ll}
%\hline
{\bf Type} &{\bf Breaking terms}   \\
\hline\\
``Gaugino'' mass $\qquad$      &$m_\la \la^\a\la_a$ \\
Scalar matter field masses $\qquad$ &$m_{ij}A^i A^j$ + conjugate \ ,   
     $\quad$  ${m'}_i{}^jA^i \AB_j$  \\
Scalar field interactions $\qquad$ &$h_{ijk}A^iA^jA^k$ + conjugate  \\
\end{tabular}
%\caption{The possible ``soft breakings'' of Ref.~\cite{gir-gri}.}
\caption{The possible ``soft breaking'' terms.}
\label{tab-soft-br}
\end{table}
%+++++++++++++++++++++++++++++++++++++++++++++++++++++++++++++++

The renormalization of the super-Yang-Mills theory including all
possible soft breakings has been carried out in~\cite{mpw2},
starting from the action \equ{sym-action}, i.e. in the
Wess-Zumino gauge.
%************************************************************
%\subsubsection*{Spontaneous breakdown of local supersymmetry:}
\remark
It has been shown~\cite{super-reviews} that precisely the 
``soft'' breaking terms listed in Table~\ref{tab-soft-br}
appear in the low energy\footnote{I.e. at energies
well below of the Planck mass.} effective theory
resulting from a supergravity theory where
{\it local} supersymmetry is spontaneously broken. There, the ``soft''
breaking terms are induced by a super-Higgs mechanism.
Thus, although a spontaneous breakdown of rigid supersymmetry seems to
be ruled out because of the apparition of a Goldstone fermion, 
the breakdown of  local supersymmetry does not present this undesirable
feature and  may lead to
phenomenologically interesting consequences.
\newpage
%******************************************************************
\subsection{Spontaneously Broken Supersymmetry}
\biblio{o'raif,ir-anom,mass-gen,ps-book}
\oilbib
The model proposed by O'Raifeartaigh~\cite{o'raif} consists of three
chiral supermultiplets
\eq
A_k = \{A_k,\p_k,F_k\}\ ,\quad k=0,1,2\ ,
\eqn{oraif-fields}
and is described by the suptab-soft-brersymmetric action
\eq\ba{l}
\S = \xint\lp \dsum{k=0}{2} \left.\AB_kA_k\right|_D  
 -\lp m\left. A_1A_2\right|_F  
 + \dfrac{g}{8} \left.A_0{A_1}^2\right|_F
            + \la F_0 + \mbox{ conj.}\rp \rp  \es
\phantom{\S} = \xint\LP \dsum{k=0}{2}\lp \pa^\m A_k\pa_\m \AB_k +
      \dfrac{i}{2}\p_k\s^\m\pa_\m\psb_k + F_k\FB_k \rp \es
\phantom{\S = \xint\LP} - m \lp A_1F_2+A_2F_1 - \half\p_1\p_2
                                      + \mbox{ conj.} \rp \es
\phantom{\S = \xint\LP} -\dfrac{g}{8} \lp {A_1}^2F_0 + 2A_0A_1F_1 
  - \half A_0\p_1\p_1 - A_1\p_0\p_1 + \mbox{ conj.} \rp  \es
\phantom{\S = \xint\LP} - \la \lp F_0 +\FB_0\rp \ . 
\ea\eqn{oraif-action}
The three parameters $m$, $g$ and $\la$ are in general complex numbers.
However they will be assumed to be real. This can indeed always be
achieved through a redefinition of the three supermultiplets by
multiplication with suitable phase factors.
Explicit expressions for the $F$ and $D$-components of the composite
superfields appearing in the first line 
have been calculated using the rules given
in Eqs.~\equ{chir-chir-prod} and \equ{chir-antichir-prod}. Recall that
the same symbol represents a supermultiplet 
as well as its first component.

The action \equ{oraif-action}
is uniquely defined by its invariances under supersymmetry,
under the discrete symmetry 
\eq
A_0'=A_0\ ,\quad A_1'=-A_1\ ,\quad A_2'=-A_2\ ,
\eqn{oraif-disc-sym}
and under the $R$-symmetry \equ{chiral-R-transf}, with particular
weights $n_k$:
\eq\ba{l}
\d_R A_k=in_kA_k\ ,\es
\d_R \p_k=i(n_k+1)\p_k\ ,\qquad n_0=n_2=-2\ ,\ n_1=0\ ,\es
\d_R F_k=i(n_k+2)F_k\ .
\ea\eqn{oraif-r-sym}
The auxiliary fields $F_k$ may be eliminated using their equations of
motion 
\eq\ba{l}
F_0(\AB) = \la + \dfrac{g}{8}{\AB_1}^2\ ,\es
F_1(\AB) = m \AB_2 + \dfrac{g}{4}\AB_0\AB_1\ ,\es
F_2(\AB) = m \AB_1\ ,
\ea\eqn{oraif-aux-f}
which leads to the potential term (c.f. \equ{baby-pot})
\eq
V(A,\AB) = \dsum{k=0}{2} \left| F(\AB)\right|^2\ .
\eqn{o'raif-pot}
Let us assume that the values of the parameters of the model are lying
in the range defined by 
\[
m^2>\dfrac{1}{8}|\la g|
\]
Then the minimum of the potential takes a positive, nonvanishing value:
\eq
\mbox{Min}(V) = \la^2\ , \quad \mbox{at:}\quad 
  A_1=A_2=0,\,A_0=a=\mbox{ arbitrary complex number.}
\eqn{oraif-min}
At the points which minimize the potential, defining the
equilibrium state of the classical theory, the auxiliary field $F_0$
takes the nonvanishing value 
\eq
F_0=\la\ .
\eqn{oraif-vev}
This characterizes a {\it
spontaneous breakdown of the supersymmetry}. Indeed, if supersymmetry
had remained intact, the equilibrium  value of any supermultiplet 
$\{A,\p,F\}$, which must be of the form $\{a,0,f\}$, with $a$ and $f$
arbitrary numbers, due to Poincar\'e invariance,
would reduce to the  value $\{a,0,0\}$, with $a$ an arbitrary number, by
consistency with the supersymmetry tranformation law
\equ{comp-chir-transf} for $\p$. But this would be in contradiction with 
the result \equ{oraif-vev}. We thus have a nonsupersymmetric equilibrium
state, although the dynamics defined by the action is symmetric: this is
just the situation of a spontaneous breakdown.

In order to see the physical consequences, let us perform the shift
\eq
F_0 \longrightarrow F_0 + \la\ ,
\eqn{oraif-shift}
such that the new field variable $F_0$ takes the value 0 in the
equilibrium state. This modifies the action, in particular,
through the appearance of an additional mass term
\[
-\dfrac{g\la}{8} \xint \lp{A_1}^2+{\AB_1}^2\rp\ ,
\]
which implies a mass splitting within the supermultiplet $A_1$, making
thus
obvious the breakdown of supersymmetry. On the other hand, the spinor
field $\p_0$ remains massless: it describes the Goldstone fermion associated
with the spontaneous breakdown of supersymmetry -- whose generator is a
fermion as we know. It is precisely this field 
which transforms
under an infinitesimal supersymmetry transformation into
the field $F_0$ with nonvanishing equilibrium value, namely (after the
shift \equ{oraif-shift}):
\[
\d_\a \p_0^\b = 2\d_\a^\b (F_0+\la)\ .
\]
It is this fact which characterizes $\p_0$ as the Goldstone fermion 
field.

%************************************************************
\subsubsection*{Renormalization and radiative mass 
generation~\cite{mass-gen,ps-book}:}
Using the scheme that we have already applied to the renormalization of 
the super-Yang-Mills theories, in Section~\ref{renormalisation}, let us
first write the BRS algebra corresponding to the supersymmetry
transformations and to the translations -- leaving aside the
$R$-transformations, which do not present any particular 
problem~\cite{ps-book}:
\eq\ba{l}
s A_k = \e^\a\p_{0\a} -i\xi^\m\pa_\m A_k,\es
s \p_{k\a} = -2\e_\a\lp F_k+\la\d_{0k}\rp
  + 2i\pa_\m A_k\s_{\a\ad}^\m\eb^\ad -i\xi^\m\pa_\m \p_{k\a} ,\es 
sF_k = 
 i\pa_\m\p_k^\a\s_{\a\ad}^\m\eb^\ad -i\xi^\m\pa_\m F_k\ ,\es
s\e^\a=s\eb^\ad=0\ ,\quad s\xi^m = -2\e^\a\s^\m_{\a\ad}\eb^\ad\ ,\es
k=0,1,2\ ,
\ea\eqn{oraif-brs}
with
\[ 
s^2=0\ .
\]
We have to show if we can implement the corresponding 
Slavnov-Taylor identity\footnote{The transformations being linear, 
here, we can write
explicitly the form of the functional operator $s$ without introducing
external fields.}  
\eq
\s\G \equiv \xint\dsum{\vf}{} s\vf\dfud{}{\vf}\G +
   s\xi^\m\dpad{}{\xi^\m}  = 0\ ,
\eqn{oraif-slavnov}
to all orders in $\h$, where 
\[
\G = \S + O(\h)
\] 
is the vertex functional or generating functional of the one-particle
irreducible Green functions (see Subsection~\ref{ren-anomalie}).

We assume \equ{oraif-slavnov} to hold up to order $n-1$, thus writing 
\[
s\G = \h^n \D + O(\h^{n+1})\ ,
\]
where $\D$ is an integrated local field polynomial of dimension $\le4$
according to the quantum action principle.
The nilpotency of $s$ again leads to the consistency condition 
\[
s\D=0\ .
\]
One finds that its most general solution is of the trivial form 
\eq
\D=s\hat\D\ ,
\eqn{oraif-delta}
where $\hat\D$ has the dimension and the quantum numbers of the
action, is $R$-invariant and could thus be absorbed  as a counterterm,
thus enforcing the Slavnov-Taylor identity to the next order. 
(C.f. Eqs.~\equ{redef-action}--\equ{slav-next-ord})

However, although this works well with almost all the terms which may
contribute to $\hat\D$, there is one single term which poses a serious
problem, namely 
\eq
\hat\D = \m^2\xint A_0\AB_0\ ,
\eqn{delta-hat} 
where $\m^2$ is a coefficient calculable as a function of the parameters 
of the theory. It turns out to be nonvanishing already in the
approximation of the one-loop graphs: 
\eq
\m^2 = \h\k \dfrac{g^4\la^2}{m^2} \ ,
\eqn{mu-square}
where $\k$ is some numerical
constant~\cite{ir-anom,ps-book}, and where we have put explicitly the
factor $\h$ in order to emphasize the quantum character of the breaking. 
As we already did in the case of the
supersymmetric gauge theories for breakings of the
form \equ{oraif-delta}, we could try to
absorb the expression $\hat\D$ as a counterterm, thus restablishing 
the Slavnov-Taylor identity. But we cannot do so as naively, since the
insertion
of this counterterm -- which is bilinear in the massless field $A_0$ --
would cut the propagator of $A_0$
\eq
\D_{A_0\AB_0}(k) = \dfrac{i}{k^2}
\eqn{azero-prop}
in two pieces, thus giving rise to a nonintegrable infrared -- i.e. 
at zero-momentum -- singularity 
$1/(k^2)^2$. It seems therefore that, not being able 
to absorb $\hat\D$, we are confronted with 
the impossibility of implementing the Slavnov-Taylor identity,
i.e. supersymmetry, for the quantum theory. This obstruction has been
named an infrared anomaly.

However, one observes\footnote{See \cite{weinberg} for an early 
reference to that matter.}  that \equ{delta-hat} has
just the form of a mass term for the 
field $A_0$~\cite{mass-gen,ps-book}. 
This means that,
although one cannot absorb it as a counterterm,  one
can nevertheles absorb it as a mass parameter -- but proportional to 
$\h$! -- in the originally massless propagator \equ{azero-prop}:
\eq
\D^{\rm new}_{A_0\AB_0}(k) = \dfrac{i}{k^2-\m^2}\ .
\eqn{azero-prop-new}
Starting the theory with this new, massive, propagator, we arrive
at a Slavnov-Taylor identity which will hold exactly in the one-loop
approximation. 

It is now possible to go on to the next orders, since the breaking 
\equ{oraif-delta} with \equ{delta-hat}, which will not keep to appear,
can now be safely absorbed as a counterterm since the propagator of
$A_0$ has become massive. There is however a pecularity, characteristic
of this phenomenon of ``mass generation'': the denominator of the
propagator containing a mass proportional to $\h$, terms in $\log\h$
will appear in the computation of the Feynman graphs. This 
means that the usual perturbation series, which is
 a series in the powers of $\h$, will be
replaced by a series in the powers of $\h$ and 
$\log\h$~\cite{genois,mass-gen,ps-book}.

\newpage
%*************************************************************
\appendix \section*{Appendix: Some Notations and Conventions}
The notations and conventions are those of~\cite{ps-book,susy96}. A more
complete set of formulas may be found in the first appendix 
of~\cite{susy96}.
\subsection*{Weyl Spinors and Pauli Matrices}%\label{spineurs-de-Weyl}
%\bi
\point{Units:} $\h=c=1$
\point{Space-time metric:} $(g_{\m\n}) = \mbox{diag}(1,-1,-1,-1)
    \ ,\quad(\m,\n,\cdots=0,1,2,3)$
\point{Fourier transform:}
\[
f(x) = \dfrac{1}{2\pi}\dint dk\;e^{ikx}\tilde f(k)\ ,\quad
\tilde f(k) = \dint dx\;e^{-ikx}f(k) \ ,\qquad
  \lp\, \pa_\m \leftrightarrow ik_\m\,\rp\ .
\]
\point{Weyl spinor:} $\lac\p_\a\ ,\ \a =1,2\rac\ 
   \in\, \mbox{repr. }(\half,0)$ of the Lorentz group. \\
  The spinor components are Grassmann variables: $\p_\a\p'_\b=-\p'_\b\p_\a$
\point{Complex conjugate spinor:} 
    $\lac\bar\p_\ad=(\p_\a)^* ,\ \ad =1,2\rac\ 
                           \in\, \mbox{repr. }(0,\half)$
\point{Raising and lowering of spinor indices:}
  \[\ba{l}
  \p^\a=\e^{\a\b}\p_\b\ ,\quad \p_\a=\e_{\a\b}\p^\b\ ,\es
  \mbox{with }   \e^{\a\b}=-\e^{\b\a}\ ,\quad\e^{12}=1\ ,\quad
  \e_{\a\b}=-\e^{a\b}\ ,\quad \e^{\a\b}\e_{\b\g}=\d^\a_\g\ ,\es
  \mbox{(the same for dotted indices).}
  \ea\]
\point{Derivative with respect to a spinor component:}
  \[\ba{l}
  \dpad{}{\p^\a}\p^\b=\d^{\b}_{\a}\ ,\quad 
     \dpad{}{\p_\a} = \e^{\a\b}\dpad{}{\p^\b}\ ,\es
  \mbox{(the same for dotted indices)}
  \ea\]
\point{Pauli matrices:}
  \[\ba{l}
  \lp\s^\m_{\a\bd}\rp=
  \lp\, \s^0_{\a\bd},\, \s^1_{\a\bd},\, 
              \s^2_{\a\bd},\, \s^3_{\a\bd}\, \rp \es
%  \phantom{\mbox{\bf Pauli matrices: }} 
  \sib_\m^{\ad\b}=\s_\m^{\b\ad}
    =\e^{\b\a}\e^{\ad\bd}\s_{\m\,\a\bd}\ ,\\[4mm]
   {(\s^{\m\n})_\a}^\b=
    \dfrac{i}{2}{\lc \s^\m\sib^\n-\s^\n\sib^\m \rc_\a}^\b   \ , \quad
  {(\sib^{\m\n})^\ad}_\bd=
    \dfrac{i}{2}{\lc \sib^\m\s^\n-\sib^\n\s^\m \rc^\ad}_\bd   \ , 
\ea\]
with
  \[\ba{l}
  \s^0=\lp\matrix{1&0\\0&1}\rp\ ,\quad  \s^1=\lp\matrix{0&1\\1&0}\rp\ ,\quad
    \s^2=\lp\matrix{0&-i\\i&0}\rp\ ,\quad   
                \s^3=\lp\matrix{1&0\\0&-1}\rp\ ,\\[5mm]
  \sib^0 =\s^0\ ,\quad \sib^i=-\s^i=\s_i\ ,\quad
  \s^{0i} = -\sib^{0i} = -i\s^i\ ,\quad 
   \s^{ij} = \sib^{ij} = \e^{ijk}\s^k\ ,\es
   \quad i,j,k=1,2,3\ .
  \ea\]
\point{Summation conventions and complex conjugation:}
Let $\p$ and $\chi$ be two Weyl spinors.
  \[\ba{l}
  \p\chi=\p^\a\chi_\a = -\chi_\a\p^\a = \chi^\a\p_\a = \chi\p \ ,\es
  \psb\chib=\psb_\ad\chib^\ad = -\chib^\ad\psb_\ad 
        = \chib_\ad\psb^\ad  = \chib\psb \ ,\es
  \p\s^\m\chib = \p^\a\smuaad\chib^\ad\ ,\quad  
        \psb\sib_\m\chi = \psb_\ad\sbmuaad\chi_\a\ ,\es
  (\p\chi)^* = \chib\psb = \psb\chib \ ,\es
  (\p\s^\m\chib)^* = \chi\s^\m\psb = -\psb\sib^\m\chi\ ,\es
  (\p\s^{\m\n}\chi)^* = \chib\sib^{\m\n}\psb\ .
\ea\]
\point{Infinitesimal Lorentz transformations of 
         the Weyl spinors:}
  \[\ba{l}
  \mbox{if}\quad  
  \d^{\rm L} x^\m = {\om^\m}_\n x^\n\ ,\quad
     \mbox{with }\om^{\m\n} = -\om^{\n\m}\qquad
      \lp\om^{\m\n}=g^{\n\rho}{\om^\m}_\rho\rp \ ,\quad\mbox{then:} 
  \es
  \d^{\rm L}\p_\a(x) = 
    \half\om^{\m\n}\lp (x_\m\pa_\n - x_\n\pa_\m)\p_\a(x)
      -\dfrac{i}{2}{(\s_{\m\n})_\a}^\b \p_\b \rp\ ,\es
  \d^{\rm L}\psb^\ad(x) = 
    \half\om^{\m\n}\lp (x_\m\pa_\n - x_\n\pa_\m)\psb^\ad(x)
      +\dfrac{i}{2}{(\sib_{\m\n})^\ad}_\bd \psb^\bd \rp\ .
  
\ea\]
%\ei

%++++++++++++++++++++++++++++++++++++++++++++++++++++++++++++
\begin{table}[hbt]
\centering
\begin{tabular}{|c||c|c|c|c|c|c|c|c|c|c|}
\hline
    &$\th^\a$ &$D_\a$ &$\f$ &$A$ &$c_+$ &$c_-$ &$B$ &$\f^*$ &$A^*$ 
     &$c_+^*$ \\
\hline\hline
$d$ &$-{1\over2}$ &${1\over2}$ &0 &1 &0 &1 &1 &2 &2 &3 \\
\hline
$n$ &$-1$ &1 &0 &$-{2\over3}$ &0 &$-2$ &$-2$ &0 &$-{4\over3}$ &-2  \\
\hline
$\F\Pi$&0 &0 &0 &0 &1 &$-1$ &0 &$-1$ &$-1$ &$-2$ \\
\hline
\end{tabular}
\caption[t1]{Dimensions  $d$, R-weights $n$ and  ghost numbers $\F\Pi$.}
\label{table1}
\end{table}
%+++++++++++++++++++++++++++++++++++++++++++++++++++++++++++++++

\newpage
%**********************************************************


\begin{thebibliography}{99}
\bibitem{gates-book} S.J. Gates, M.T. Grisaru,
%  M. Ro$\check {\rm c}$ek and W. Siegel, 
  M. Ro{\v c}ek and W. Siegel, 
  ``Superspace: or One Thousand and One Lessons 
   in Supersymmetry'', {\em (Benjamin/Cummings, London, 1983)};
\bibitem{bagger-book} J. Bagger and  J. Wess, 
  ``Supersymmetry and  Supergravity'', 
   {\em (Princeton University Press, 1983)};
\bibitem{srivatsana} P.P. Srivastava, ``Supersymmetry, 
  Superfields and Supergravity: An Introduction'',
  {\em (Bristol, Uk: Hilger, 1986)};
\bibitem{west-book} P. West, ``Introduction to Supersymmetry and
   Supergravity'', {\em (World Scintific, Singapore, 1987)};
\bibitem{buchbinder} I.L. Buchbinder and S.M. Kuzenko, 
   ``Ideas and Methods of Supersymmetry and Supergravity'',
   {\em (Institute of Physics Publishing, Bristol and Philadelphia, 1995)};
\bibitem{ps-book} O. Piguet and K. Sibold,
                         {``Renormalized Supersymmetry''},
                       series {\em ``Progress in Physics'', vol. 12
                         (Birkh\"auser Boston Inc., 1986)};
\bibitem{sohnius} M. Sohnius, ``Supersymmetry for beginners'', in 
``Supersymmetry and Supergravity'', p.3, edited by S. Ferrara et al., 
{\em World Scientific, 1983};
\bibitem{lykken} J.D. Lykken, ``Introduction to supersymmetry'', 
{\em hep-th}/9612114;
\bibitem{reviews} M. Jacob, ``Supersymmetry and Supergravity'' (reprints
from {\em Physics Reports}),    {\em (North Holland, Amsterdam, 1987};
\bibitem{susy96} O. Piguet, {``Supersymmetry, Supercurrent
and Scale Invariance''}, Lecture notes, CBPF preprint, {\em hep-th/}9611003; 
\bibitem{harm-s-space} E.A. Ivanov, ``Harmonic Superspace: New
  Directions'', in:  ``Problems of  Quantum Field Theory'', p.374
 {\em (Ed.: D.V.
 Shirkov, D.I. Kazakov \& A.A. Vladimirov,  Dubna, 1996, hep-th/}9609090;\\
E. Sokatchev, ``Harmonic Superspace and its Applications in Extended
  Supersymmetry'', in the proceedings of ``Supersymmetry and its
  Applications'', p.283,  {\em Cambridge, 1985};
\bibitem{susy-harm-osc} E. Witten, \np{B185}{81}{513}\\
M. de Crombrugghe and V. Rittenberg,   \annp{151}{83}{99}\\
F. Cooper, \prep{251}{95}{267}\\
B. de Wit, ``Supersymmetric quantum mechanics, supermembranes and 
Dirichlet particles'', Invited talk given at the $30^{\rm th}$ 
International Symposium Ahrenshoop on the Theory of Elementary 
Particles, Buckow, 1996; to appear in Nuclear 
Physics B (Proc. Suppl.), {\em hep-th}/9701169;
\bibitem{col-mand} S. Coleman and J. Mandula, \pr{159}{67}{1251}
\bibitem{haag-lo-so} R. Haag, J. {\L}opusza\'nski 
     and M. Sohnius, \pl{B88}{75}{257}
\bibitem{sym-theories} A. Salam and J. Strathdee, \pl{B51}{74}{353}\\
S. Ferrara and B. Zumino, \np{B79}{74}{413}
\bibitem{bm}     P. Breitenlohner and D. Maison,
                 ``Renormalization of supersymmetric Yang--Mills
                 theories'', {\em
                 in Cambridge 1985, Proceedings:
                 ``Supersymmetry and its applications''},
                 p.309;\\
                 and ``N=2 Supersymmetric Yang--Mills theories
                 in the Wess--Zumino gauge'', {\em in
                 ``Renormalization of quantum field
                 theories with
                 nonlinear field transformations'',
                 ed.: P. Breitenlohner, D. Maison and K.
                   Sibold,
                 Lecture Notes in Physics, Vol.} 303,
                  p.64 ({\em Springer Verlag, Berlin,
                 Heidelberg}, 1988);
\bibitem{white2} P.L. White, \cqg {9}{92}{413}
\bibitem{white1} P.L. White, \cqg {9}{92}{1663}
\bibitem{maggiore}  N. Maggiore, \ijmp{A10}{95}{3937,3781}
%   ``Off-shell formulation of N=2 Super Yang--Mills theories coupled 
%     to matter without auxiliary fields'',  hep-th/9412092,
%   ``Algebraic renormalization of N=2 Super Yang--Mills theories 
%     coupled to matter'', hep-th/9501057,
\bibitem{mpw1} N. Maggiore, O. Piguet and S. Wolf, \np{B458}{96}{403}
% Algebraic Renormalization of $N=1$ Supersymmetric Gauge Theories,
\bibitem{mpw2} N. Maggiore, O. Piguet and S. Wolf, \np{B476}{96}{329}
% Algebraic Renormalization of $N=1$ Supersymmetric Gauge Theories with 
%Supersymmetry Breaking masses,
\bibitem{sohnius-west}   M.F. Sohnius and P.C. West,
                                \pl{B100}{81}{245} \\
  P.C. West, ``Supersymmetry and Finiteness'', 
   {\em  Shelter Island II} 1983:127;
\bibitem{mandelstam-brink} S. Mandelstam, \np{B213}{83}{149}\\
  L. Brink, O. Lindgren and B. Nilsson, \np{B212}{83}{401}
\bibitem{super-reviews} H.P. Nilles, \prep{110}{84}{1} \\
%  ``Supersymmetry,   Supergravity and   Particle Physics'', 
H.E. Haber and G.L. Kane, \prep{117}{85}{75}\\
R.N. Mohapatra, ``Unification and Supersymmetry'', 
  ({\em Springer-Verlag, New York, 1986}); 
\bibitem{p-sor-book} O. Piguet and S.P. Sorella, ``Algebraic
                 Renormalization'', {\em Lecture Notes in Physics,
                 Vol.} m28
                 ({\em Springer Verlag, Berlin,
                 Heidelberg}, 1995); 
\bibitem{sorella-vitoria} F. Fucito, A. Tanzini, L.C.Q. Vilar, 
  O.S. Ventura, C. Sasaki, V. Lemes and S.P. Sorella, 
  ``Lectures on Algebraic Renormalization'', {\em in these proceedings};
\bibitem{bbbcd} C. Becchi, A. Blasi, G. Bonneau, R. Collina and F.
                 Delduc,\\ \cmp{120}{88}{121}
\bibitem{dixon}  J.A. Dixon, \cmp {140}{91}{169}
\bibitem{chiral-eff-th} H. Leutwyler, ``Foundations and Scope of Chiral
 Perturbation Theory'', in ``Chiral Dynamics Workshop 1994'', p.14, 
{\em hep-ph}/9409423;\\
J. Bijnen, G. Ecker and J. Gasser, ``Chiral Perturbation Theory'', 
   {\em hep-ph}/9411232;
\bibitem{o'raif} L. O'Raifeartaigh, \np{B96}{75}{331}
\bibitem{gir-gri} L. Girardello and M.T. Grisaru, \np{B194}{82}{65}
\bibitem{ir-anom} O. Piguet and K. Sibold, \np{B119}{77}{292}
\bibitem{mass-gen} W. Bardeen, O. Piguet and K. Sibold,
                  \pl{B72}{77}231\\    
O. Piguet, M. Schweda and K. Sibold, \np{B168}{80}{337}
\bibitem{weinberg} S. Weinberg, \pr{D7}{73}{2887}
\bibitem{genois} G. Bandelloni, C. Becchi, A. Blasi and R.Collina, 
   \cmp{67}{79}{147}
\end{thebibliography}
\end{document}